\documentclass[twocolumn,prd,aps,nofootinbib,showpacs,superscriptaddress]{revtex4-1}
\usepackage{longtable}
\usepackage{multirow}
\usepackage{graphicx}
\usepackage[dvips]{color}
\usepackage{amssymb,amsmath}
\usepackage{supertabular}
\allowdisplaybreaks[4]
\begin{document}

\title{$I=2$ $\pi\pi$ $s$-wave scattering length from lattice QCD}

\author{Ziwen Fu}
\email{fuziwen@scu.edu.cn}
\affiliation{
Key Laboratory of Radiation Physics and Technology of Education Ministry;
Institute of Nuclear Science and Technology, Sichuan University, Chengdu 610064, P.R. China
}

\affiliation{
The Center for Theoretical Physics, College of Physics, Sichuan University,
Chengdu 610064, P.R. China
}

\author{Wang  Jun}
\email{wangjun@scu.edu.cn}
\affiliation{
Key Laboratory of Radiation Physics and Technology of  Education Ministry;
Institute of Nuclear Science and Technology, Sichuan University, Chengdu 610064, P.R. China
}


\begin{abstract}
The $I=2$ $\pi\pi$ elastic $s$-wave scattering phase shift is measured by lattice QCD with $N_f=3$ flavors of
the Asqtad-improved staggered fermions. The lattice-calculated energy-eigenvalues of $\pi\pi$ systems
at one center of mass frame and some moving frames using the moving wall source technique are utilized
to secure phase shifts by L\"uscher's formula.
Our computations are fine enough to obtain threshold parameters:
scattering length $a$, effective range $r$, and shape parameter $P$,
which can be extrapolated at the physical point by NLO in chiral perturbation theory,
and our relevant NNLO predictions from expanding NPLQCD's works are novelly
considered as the systematic uncertainties.
Our outcomes are consistent with Roy equation determinations, newer experimental data, and lattice estimations.
Numerical computations are performed with a coarse
($a\approx0.12$~fm, $L^3 T = 32^3 64$),
two fine ($a\approx0.09$~fm, $L^3 T = 40^3 96$)
and a superfine ($a\approx0.06$~fm, $L^3 T = 48^3 144$) lattice ensembles
at four pion masses of $m_\pi\sim247~{\rm MeV}$, $249~{\rm MeV}$, $275~{\rm MeV}$,
and $384~{\rm MeV}$, respectively.
\end{abstract}

\maketitle

\section{Introduction}
Pion-pion ($\pi\pi$)  scattering at low energies is the basic 
and most-understood hadronic scattering processes.
Its scattering amplitudes can be uniquely predicted at leading order (LO)
in chiral perturbation theory($\chi$PT)~\cite{Weinberg:1966kf}.
The next-to-leading order (NLO) and next-to-next-to-leading order (NNLO) chiral expansions
lead to minor deviations from LO prediction for small pion masses,
and involve the computable non-analytical contributions and analytical terms with low-energy
constants(LEC's)~\cite{Gasser:1983yg,Bijnens:1995yn,Bijnens:1997vq,Colangelo:2001df,Bijnens:2014lea},
which can be directly achieved from both experimental measurements and lattice computations.

The E865 obtains $\pi\pi$ scattering lengths from
the semileptonic $K_{e4}$ decay using strict $\chi$PT constraints~\cite{Pislak:2003sv}.
Using Roy equations, the NA48/2 analyses of the $K_{e4}$ and
$K_{3\pi}$ decays lead to the robust results
on the $s$-wave $\pi\pi$ scattering lengths~\cite{Batley:2010zza}.
All of these numbers can be employed to inversely secure the LEC's values.

Lattice calculations of $I=2$ $\pi\pi$ scattering have been investigated
by various lattice groups~\cite{Sharpe:1992pp,Kuramashi:1993ka,Fukugita:1994ve,
CP-PACS:2002wru,CP-PACS:2004dtj,Beane:2005rj,Beane:2007xs,Sasaki:2008sv,Feng:2009ij,Fischer:2020jzp,Dudek:2010ew,Yagi:2011jn,NPLQCD:2011htk,Dudek:2012gj,
Fu:2013ffa,Kurth:2013tua,Sasaki:2013vxa,Helmes:2015gla,Bulava:2016mks,HALQCD:2017xsa,Culver:2019qtx,RBC:2021acc,Rodas:2023gma},
which are found to be weak and repulsive, as in experiment~\cite{Pislak:2003sv,Batley:2010zza},
and the scattering length $a$ can be measured with a noteworthy level of precision.
It has been customary to use a standard effective range expansion parametrization (ERE)
for the isospin-2 $\pi\pi$ scattering~\cite{NPLQCD:2011htk,Dudek:2012gj,Sasaki:2013vxa,Helmes:2015gla,Fischer:2020jzp},
and at the present time, the relevant effective range $r$ can be also estimated with high
precision~\cite{NPLQCD:2011htk,Dudek:2012gj,Sasaki:2013vxa,Helmes:2015gla,Fischer:2020jzp}.
Furthermore, the variation of the scattering length with changing quark mass has been explored
under the guidance of the chiral perturbation theory~\cite{NPLQCD:2011htk,Fischer:2020jzp}.

With tremendously improved computing capabilities, lattice simulations can calculate LEC's values in
$I=2$ $\pi\pi$ scattering with robust statistics.
Using LEC's obtained at nonphysical pion masses,
threshold parameters and effective range expansion parameters
at physical point can be extrapolated in $\chi$PT.
Moreover, Roy-equation~\cite{Roy:1971tc,Basdevant:1973ru,Ananthanarayan:2000ht}
can determine $\pi\pi$ scattering parameters
with trustworthy precision~\cite{Bijnens:1997vq,Colangelo:2001df,GarciaMartin:2011cn},
which can be used to compare with relevant lattice evaluations.
NPLQCD have elegantly exhibited these strategies~\cite{NPLQCD:2011htk}.

It is very impressive that, as illustrated for $I=2~\pi\pi$ scattering~\cite{NPLQCD:2011htk},
near threshold behavior of inverse partial wave amplitude can be used to get scattering length,
effective range and shape parameters, which can be written in terms of a few LEC's
that meet with low-energy theorems mandated by chiral symmetry,
and encrypt the chief momentum dependence of partial wave amplitude~\cite{Bijnens:1997vq,NPLQCD:2011htk}.
Thus, one can predict the scattering parameters at physical point
by NLO $\chi$PT expressions,
which are expressed in terms of three independent LEC's~\cite{NPLQCD:2011htk}.

We extend NPLQCD's skill to $I=0$ $\pi\pi$ scattering~\cite{Fu:2017apw}.
The NLO expressions for the effective range and shape parameter
are found to be valid for range of interest~\cite{Fu:2017apw},
and more higher order terms should be added for more sophisticated study~\cite{Fu:2017apw}.
Moreover, we realize that high order terms
actually contain the information from  both NLO and NNLO expressions in $\chi$PT.
This motives us to deeply study $\pi\pi$ scattering at NNLO.

Since $I=2$ $\pi\pi$ scattering is technically the easiest case to
study in lattice QCD, due to the absence of the disconnected contributions,
its phase shifts can be precisely gotten,
and it is best-choice to look into the validity and scope of the relevant NNLO expressions,
which are regarded as an important systematic uncertainty for the relevant NLO expressions~\cite{NPLQCD:2011htk}.

This work is also inspired and strongly assisted by the Yagi's pioneering work~\cite{Yagi:2011jn},
where $I=2$ $\pi\pi$ scattering length is expanded at NNLO by the chiral expansion parameter $\xi$
with three LEC's~\cite{Yagi:2011jn}, and similarly for the slope parameter $b$~\cite{Bijnens:1997vq}.
We exploringly pick up the slope parameter $c$ at NNLO 
from $\pi\pi$ scattering amplitude~\cite{Bijnens:1997vq}, and rearrange it with Yagi's strategies~\cite{Yagi:2011jn},
which leads to another three LEC's.
Consequently, six more LEC's are needed to expand NPLQCD's skill~\cite{NPLQCD:2011htk}
to analyze $I=2$ $\pi\pi$ scattering amplitude at NNLO in $\chi$PT.

Meanwhile, we partially confirmed from perspective of theoretical analysis and phenomenological values
that relevant NLO expressions in Ref.~\cite{NPLQCD:2011htk}
are fully compatible with relevant NNLO expressions and phenomenological values
for relatively small pion mass ($400\rm MeV$)~\cite{Bijnens:1997vq,Colangelo:2001df,GarciaMartin:2011cn}.
Admittedly, it is a little surprising that both methods's relevant
results are pretty accordant within statistical uncertainties.

As demonstrated later, although it is enough to use NLO $\chi$PT expressions to fit the present lattice data,
it is not a trivial extension since it predicts more reasonable and meaningful outcomes as compared with NLO's,
especially for relatively larger pion mass, 
it proposes a helpful approach towards more sophisticated comprehension and computation
of $\pi\pi$ scattering with lattice QCD for future more accurate lattice study.

It must be acknowledged that the enlightening work in Ref.~\cite{Bulava:2016mks}
is beneficial to choose some lattice irreducible representations (irreps),
which can produce small scattering momenta $k$,
and it turns out to be important to fairly secure the proper effective range expansion parameters.

It's best to understand that these methods are just effective in the elastic region~\cite{Adhikari:1983ii,Beane:2003da}.
Although the strict value of this threshold is not clear,
the threshold $m_\pi< 400\, \rm MeV$ should be reasonable choice~\cite{Hanhart:2008mx,Hanhart:2014ssa,Albaladejo:2012te,Pelaez:2010fj}.

In this work, we exploit one MILC coarse ($a\approx0.12$~fm, $L^3 \times T = 32^3\times 64$),
two fine ($a\approx0.09$~fm, $L^3 \times T = 40^3\times 96$),
and one superfine ($a\approx0.06$~fm, $L^3 \times T = 48^3\times 144$)
lattice ensembles with $N_f = 3$ flavors of Asqtad-improved staggered
dynamical quarks~\cite{Aubin:2004wf,Bernard:2001av,stag_fermion}
to compute $s$-wave $I=2$ $\pi\pi$ phase shift,
where L\"uscher's technique~\cite{Luscher:1986pf,Luscher:1990ux,Luscher:1990ck} and its extensions~\cite{Beane:2003da,Rummukainen:1995vs,Kim:2005gf,Christ:2005gi,Doring:2012eu,
Fu:2011xz,Leskovec:2012gb,Doring:2012eu} are employed to get
scattering phases with lattice-calculated energy eigenstates.
The moving wall source technique~\cite{Kuramashi:1993ka,Fukugita:1994ve}
is utilized to calculate the relevant quark-line diagrams~\cite{Sharpe:1992pp,Kuramashi:1993ka,Fukugita:1994ve}.

From the discussions in Refs.~\cite{Lepage:1989hd,Fu:2016itp},
if we use the finer gauge configurations,
employ lattice ensembles with relatively large spatial dimensions $L$,
and sum the correlators over all time slices ($64$, $96$ or $144$),
the signals are anticipated to be significantly improved~\cite{Fu:2016itp}.
It allows us to not only measure scattering length, but also examine the effective range and shape parameter.
The chiral extrapolation of the scattering length $m_\pi a_{0}^{I=2}$ is
carried out using NLO $\chi$PT. Extrapolated to the physical point, 
our final results give rise to
$$
m_\pi a_{0}^{I=2} = -0.04433(32)(163),
$$
where the systematic error is estimated from the fitting
with relevant NNLO representations~\cite{Bijnens:1997vq,Colangelo:2001df},
as suggested by NPLQCD~\cite{NPLQCD:2011htk},
and it is in reasonable agreement with the recent experimental and theoretical
determinations as well as the corresponding lattice calculations.

Most of all, after chiral extrapolations of effective range $m_\pi r$ and shape parameter
to physical point, we obtain
\begin{eqnarray}
m_\pi r &=&  57.41(1.04)(4.76), \cr
P       &=& -52.79(1.09)(8.66), \nonumber
\end{eqnarray}
where the systematic uncertainties are also estimated with our derived NNLO representations~\cite{NPLQCD:2011htk}.
They are in fair accordance with the Roy-equation determinations~\cite{Bijnens:1997vq,Colangelo:2001df}
and some relevant lattice studies~\cite{NPLQCD:2011htk,Dudek:2012gj,Sasaki:2013vxa,Helmes:2015gla,Fischer:2020jzp}.

This paper is arranged as follows.
L\"uscher's method and lattice scheme are discussed in Sec.~\ref{Sec:Methods}.
Lattice results are given in Sec.~\ref{sec:pipiscattering},
along with relational fits, which are used to gain threshold parameters.
A derivation of the relevant $\chi$PT formulas at NNLO
and the chiral extrapolation of lattice-measured data are presented in Sec.~\ref{sec:chiextrap}.
A simple summary and discussion are shown in Sec.~\ref{sec:discussion}.
The derivation of the scattering amplitude at NNLO in $\chi$PT is courteously dedicated to the Appendix~\ref{app:ChPT NNLO}.
The discussion of NNLO $\chi$PT expressions is provided in Appendix~\ref{app:C NNLO}.
The scale dependence of coupling constant $\tilde{r}_i(i=1-6)$ is explicitly offered
in Appendix~\ref{app:R NNLO}.

%
\section{Finite-volume methods}
\label{Sec:Methods}
In the present study, we will examine the $s$-wave $\pi\pi$ system with
the isospin representation of $(I,I_z)=(2,2)$.
The calculations are carried out for total
momenta $\mathbf{P}=[0,0,0]$, $[0,0,1]$, $[0,1,1]$, $[1,1,1]$, $[0,0,2]$,
 $[0,2,2]$, $[0,0,3]$, $[2,2,2]$, and $[0,0,4]$,
which are written in units of $2\pi/L$.

\subsection{Center of mass frame}
In the center-of-mass (CoM) frame, the energy levels of two free pions are provided by
$$
E = 2\sqrt{m_\pi^2+ |{\mathbf p}|^2} ,
$$
where ${\mathbf p}=\tfrac{2\pi}{L}{\mathbf n}$, and ${\mathbf n}\in \mathbb{Z}^3$.
The lowest energy $E$ for ${\mathbf n} \ne 0$
is beyond the $t$-channel cut, which starts at ${k^2}={m_\pi^2}$~\cite{NPLQCD:2011htk}.

The energy levels of $\pi\pi$ system are shifted
by the hadronic interaction from $E$ to $\overline{E}$,
$$
\overline{E} = 2\sqrt{m_\pi^2 + k^2} , \quad k=\frac{2\pi}{L}q ,
$$
where the dimensionless scattering momentum $q \in \mathbb{R}$.
It is the L\"uscher formula that relates the energy $\overline{E}$
to the $s$-wave $\pi\pi$ scattering phase
$\delta$~\cite{Luscher:1990ux,Luscher:1990ck},
\begin{equation}
\label{eq:CMF}
k \cot\delta(k)=\frac{2}{L\sqrt{\pi}} {\mathcal{Z}_{00}(1;q^2)} ,
\end{equation}
where the zeta function is formally defined by
\begin{equation}
\label{eq:Zeta00_CM}
\mathcal{Z}_{00}(s;q^2)=\frac{1}{\sqrt{4\pi}}
\sum_{{\mathbf n}\in\mathbb{Z}^3} \frac{1}{\left(|{\mathbf n}|^2-q^2\right)^s}  .
\end{equation}
The zeta function $\mathcal{Z}_{00}(s;q^2)$
can be efficiently evaluated by the method described in Refs.~\cite{Yamazaki:2004qb,Doring:2011vk}.

\subsection{Moving frame}
Using a moving frame with non-zero total momentum
${\mathbf P}=\frac{2\pi}{L}{\mathbf d}, {\mathbf d}\in\mathbb{Z}^3$,
energy level of two  pions is shown as
$$
E_{MF} = \sqrt{m_\pi^2 + |{\mathbf p}_1|^2} +
         \sqrt{m_\pi^2 + |{\mathbf p}_2|^2} ,
$$
where ${\mathbf p}_1$, ${\mathbf p}_2$ denote three-momenta of two pions,
which observe the periodic boundary condition,
${\mathbf p}_1=\frac{2\pi}{L}{\mathbf n}_1$,
${\mathbf p}_2=\frac{2\pi}{L}{\mathbf n}_2$,
${\mathbf n}_1,{\mathbf n}_2\in \mathbb{Z}^3$,
and ${\mathbf P}$ satisfies
${\mathbf P} = {\mathbf p}_1 + {\mathbf p}_2$~\cite{Rummukainen:1995vs}.
The energy $E_{CM}$  is
$$
E_{CM} = 2\sqrt{m_\pi^2 + k^{2}} , \quad k = \frac{2\pi}{L} q ,
$$
where $q \in \mathbb{R}$, $k =| {\mathbf p}|$ and ${\mathbf p}$ are quantized to the values
$
{\mathbf p} =\frac{2\pi}{L}{\mathbf r}, {\mathbf r} \in P_{\mathbf d} ,
$
and the set $P_{\mathbf d}$  is
\begin{equation}
\label{eq:set_Pd_MF}
P_{\mathbf d} = \left\{ {\mathbf r} \left|  {\mathbf r} = \vec{\gamma}^{-1}
\left[{\mathbf n}+ \frac{\mathbf d}{2} \right], \right.  {\mathbf n}\in\mathbb{Z}^3 \right\} ,
\end{equation}
where $\vec{\gamma}^{-1}$ is inverse Lorentz transformation
acting in direction of  CoM velocity ${\mathbf v}$,
$
\vec{\gamma}^{-1}{\mathbf p} =
\gamma^{-1}{\mathbf p}_{\parallel}+{\mathbf p}_{\perp} ,
$
where ${\mathbf p}_{\parallel}$ and ${\mathbf p}_{\perp}$ are
components of ${\mathbf p}$ parallel
and perpendicular to ${\mathbf v}$, respectively.
The energy $E_{CM}$ is linked to $E_{MF}$  with the relationship
$E_{CM}^2 = E_{MF}^2-{\mathbf P}^2.$

To get more eigenenergies, we implement eight move frame~(MF),
i.e., first moving frame (MF1) are taken with ${\mathbf d}={\mathbf e}_3$,
MF2 with ${\mathbf d}={\mathbf e}_2+{\mathbf e}_3$,
MF3 with ${\mathbf d}={\mathbf e}_1+{\mathbf e}_2 + {\mathbf e}_3$,
MF4 with ${\mathbf d}=2{\mathbf e}_3$,
MF5 with ${\mathbf d}=2{\mathbf e}_2 + 2{\mathbf e}_3$,
MF6 with ${\mathbf d}=3{\mathbf e}_3$,
MF7 with ${\mathbf d}={2\mathbf e}_1+2{\mathbf e}_2 + 2{\mathbf e}_3$, and
MF8 with ${\mathbf d}={4\mathbf e}_3$.
For MF2 and MF4, the ground and first excited states can be considered.
The scattering phase shifts can be gained from total energy of two-particle system enveloped in a cubic torus
by L\"uscher technique~\cite{Luscher:1986pf,Luscher:1990ux,Luscher:1990ck,Rummukainen:1995vs}
\begin{equation}
k \cot\delta(k)  =  \frac{2}{\gamma L \sqrt{\pi}} Z_{00}^{\bf d}(1; q^{2})\,,
\label{eqn:RumGott}
\end{equation}
where $\gamma=E_{MF}/E_{CM}$.
The computation of zeta functions $\mathcal{Z}_{00}^{{\mathbf d}} (1; q^2)$
is discoursed in Ref.~\cite{Yamazaki:2004qb}.

\subsection{$\pi\pi$ correlator function}
 The isospin-$2$ $\pi\pi$ state with momentum ${\mathbf P} = {\mathbf p}_1 + {\mathbf p}_2$
 is built with following interpolating operator~\cite{Sharpe:1992pp,Kuramashi:1993ka,Fukugita:1994ve}.
$$
{\cal O}_{\pi\pi}^{I=2} (\mathbf{P},t) = \pi^{+}(\mathbf{p}_1,t) \pi^{+}(\mathbf{p}_2,t+1)
$$
with the interpolating pion operators denoted by~\cite{Sharpe:1992pp,Kuramashi:1993ka,Fukugita:1994ve}.
$$
{\pi^+}(t) = -\sum_{\bf{x}} \bar{d}({\bf{x}}, t)\gamma_5 u({\bf{x}},t) .
$$
Note that the operator ${\cal O}_{\pi\pi}^{I=2} (\mathbf{P},t)$ belongs to $A_1^+$~\cite{Bulava:2016mks}.

The $\pi\pi$ four-point correlation function
with the total momentum ${\mathbf P}$ can be expressed as,
\begin{eqnarray}
\label{EQ:4point_pK_mom}
&&\hspace{-0.3cm}C_{\pi\pi}({\mathbf P}, t_4,t_3,t_2,t_1) =
\sum_{\mathbf{x}_1,\mathbf{x}_3}\sum_{\mathbf{x}_2,\mathbf{x}_4}
e^{ i{{\mathbf p}_1} \cdot ({\mathbf{x}}_3 -{\mathbf{x}}_1) } e^{ i{{\mathbf p}_2} \cdot ({\mathbf{x}}_4 -{\mathbf{x}}_2) }\cr
&&\hspace{0.80cm}\times\langle
{\cal O}_{\pi}({\bf{x}}_4, t_4)
{\cal O}_{\pi}({\bf{x}}_3, t_3)
{\cal O}_{\pi}^{\dag}({\bf{x}}_2, t_2)
{\cal O}_{\pi}^{\dag}({\bf{x}}_1, t_1) \rangle ,
\end{eqnarray}
where ${\mathbf P} = {\mathbf p}_1 + {\mathbf p}_2$,
and one typically opts $t_1=0$, $t_2=1$, $t_3=t$, and $t_4=t+1$ to hinder color Fierz rearrangement
of quark lines~\cite{Kuramashi:1993ka,Fukugita:1994ve}.

In the isospin limit, $I=2$ $\pi\pi$ scattering amplitude just contains two components,
and their quark-line diagrams are demonstrated in Fig.~1 of the former works~\cite{Fu:2013ffa,Fu:2017apw},
which are categorized as the direct ($D$) and crossed ($C$) diagrams~\cite{Kuramashi:1993ka,Fukugita:1994ve,Sharpe:1992pp}.

The moving wall source technique is usually exploited
to compute quark-line diagrams~\cite{Kuramashi:1993ka,Fukugita:1994ve}.
In our previous works~\cite{Fu:2012gf,Fu:2013ffa},
a detailed procedure is devoted to express
these diagrams in CoM frame~\cite{Fu:2013ffa}
with the light quark propagator $G$~\cite{Kuramashi:1993ka,Fukugita:1994ve},
and the relevant  representations in the moving frame are analogically offered in Ref.~\cite{Fu:2012gf}.
To deliver them in the generic frame,
we employ an up quark source with $1$, and an
anti-up quark source with $e^{i{\mathbf p} \cdot {\mathbf{x}} }$
on each site for two pion creation operator, respectively, then relevant expressions
can be expressed in terms of the light $u/d$ quark propagator $G$
as~\cite{Kuramashi:1993ka,Fukugita:1994ve,Fu:2012gf,Fu:2013ffa}
\begin{widetext}
\begin{eqnarray}
\label{eq:dcr}
C^D_{\pi\pi}({\mathbf P},t_4,t_3,t_2,t_1) &=&
\sum_{ \mathbf{x}_3} \sum_{ \mathbf{x}_4}
e^{ i{{\mathbf p}_1} \cdot {\mathbf{x}}_3 } e^{ i{{\mathbf p}_2} \cdot {\mathbf{x}}_4 }
\langle \mbox{Tr}
[G_{t_1}^{\dag}({\mathbf{x}}_3,t_3)G_{t_1}({\mathbf{x}}_3,t_3)]
\mbox{Tr}
[G_{t_2}^{\dag}({\mathbf{x}}_4,t_4)G_{t_2}({\mathbf{x}}_4,t_4)] \rangle,\cr
C^C_{\pi\pi}({\mathbf P},t_4,t_3,t_2,t_1) &=&
\sum_{ \mathbf{x}_3} \sum_{ \mathbf{x}_4}
e^{ i{{\mathbf p}_1} \cdot {\mathbf{x}}_3 } e^{ i{{\mathbf p}_2} \cdot {\mathbf{x}}_4 }
\langle \mbox{Tr}
[G_{t_1}^{\dag}({\mathbf{x}}_3,t_3)G_{t_2}({\mathbf{x}}_3,t_3)
 G_{t_2}^{\dag}({\mathbf{x}}_4,t_4)G_{t_1}({\mathbf{x}}_4,t_4)] \rangle,
\end{eqnarray}
\end{widetext}
where ${\mathbf P} = {\mathbf p}_1 + {\mathbf p}_2$.
The combinations of the light quark propagators $G$ for quark-line diagrams
that are adopted for the $\pi\pi$ correlation
functions are illustrated in Fig.~1 of Refs.~\cite{Fu:2013ffa,Fu:2017apw}.
The corresponding $I=2$  $\pi\pi$ scattering correlation functions
can be delivered in terms of  just two quark-line diagrams~\cite{Sharpe:1992pp,Kuramashi:1993ka,Fukugita:1994ve},
$$
C_{\pi\pi}^{I=2}(t) \equiv D - N_f C ,
$$
where $N_f=4$ is staggered-flavor factor due to the number of tastes intrinsic
to Kogut-Susskind formulation~\cite{Sharpe:1992pp}.
\begin{table*}[t!]
\caption{ \label{tab:MILC_configs}
Simulation parameters of MILC lattice ensembles.
Lattice dimensions are written in lattice units with spatial ($L$) and temporal ($T$) size.
Column $3$ indicates gauge coupling $\beta$.
The fourth block gives bare masses of light and strange quark masses.
Column $5$ gives pion masses in MeV.
Lattice spatial dimension in units of pion mass and ratio $m_\pi/f_\pi$
are given in Column $6$ and $7$ respectively.
The number of gauge configurations and the number of time slices calculated light propagators are shown in last Column.
}
\begin{ruledtabular}
\begin{tabular}{llllccll}
Ensemble &$L^3 \times T$ &$\beta$ & $a m_l/a m_s$ & $m_\pi({\rm MeV})$ &$m_\pi L$  & $m_\pi/f_\pi$
 &$N_{\rm cfg} \times N_{\rm slice}^{\pi\pi}$   \\
\hline
\multicolumn {8}{c}{$a \approx 0.06$~fm}        \\
48144f21b7475m0054m018 & $48^3\times144$ & $7.475$   & $0.0054/0.018$  & $384$  & $5.48$ & $2.57(2)$  &  $174\times 144$  \\
\multicolumn {8}{c}{$a \approx 0.09$~fm}        \\

4096f21b708m0031m031  & $40^3\times96$   & $7.08$    & $0.0031/0.031$  & $247$  & $4.21$  & $1.77(1)$ & $604\times 96$ \\
4096f3b7045m0031      & $40^3\times96$   & $7.045$   & $0.0031/0.0031$ & $249$  & $4.20$  & $1.79(1)$ & $560\times 96$ \\
\multicolumn {8}{c}{$a \approx 0.12$~fm}        \\
3264f3b6715m005       & $32^3\times64$   & $6.715$   & $0.005/0.005$   & $275$  & $5.15$  & $1.95(2)$ & $696\times 64$
\end{tabular}
\end{ruledtabular}
\end{table*}

\subsection{Lattice Calculation}
We employ gauge configurations with three Asqtad-improved
staggered sea quarks~\cite{Kaplan:1992bt,Bernard:2001av,Bernard:2010fr,Bazavov:2009bb}.
The simulation parameters are listed in Table~\ref{tab:MILC_configs}.~\footnote{
We thank professor Carleton DeTar kindly give us some MILC lattice ensembles for our initial lattice study.
More lattice gauge configurations can be obtained by updating the current gauge configurations
to produce fresh trajectories or directly generate new ones with the help of MILC codes~\cite{MILC:DeTar}.
}
As a MILC convention, it is proper to adopt $(am_l, am_s)$ to classify lattice ensembles.
Lattice ensembles are gauge fixed to Coulomb gauge
before computing quark propagators.

Despite it is expensive, moving wall source technique
can be used to calculate relevant correlators with high quality~\cite{Kuramashi:1993ka,Fukugita:1994ve}.
This method is widely extended to two-particle system with nonzero momenta~\cite{Fu:2011xw,Fu:2012gf,Fu:2012tj,Fu:2013ffa,Fu:2011wc,Fu:2017apw}.
From these practical computations,
moving-wall source technique is identified  to calculate
four-point correlators with desirable quality.

The correlators are determined on all the $T$ time slices,
namely, the correlator $C_{\pi\pi}(t)$ is estimated by
\begin{eqnarray}
 C_{\pi\pi}(t) &=&
\frac{1}{T}\sum_{t_s=0}^{T-1} \left\langle
\left(\pi\pi\right)(t+t_s)\left(\pi\pi\right)^\dag(t_s)\right\rangle .
\nonumber
\end{eqnarray}
After averaging the propagators over all the $T$ values,
the statistics are anticipated to be much improved.~\footnote{
For each configuration, we measure $3T$ $u$ quark propagators ($192$, $284$, or $432$).
In practice, all the $u$ quark propagators are saved into moving disks,
and copied into memory when needed.
Using this trick, then for each time-slice computation,
it averagely costs just one light quark propagator calculation for a color.
From this perspective, it is actually ``economical''~\cite{Fu:2017apw}.
}

According to the arguments~\cite{Lepage:1989hd,Fukugita:1994ve,Fu:2016itp},
noise-to-signal ratio of $\pi\pi$ correlator
is improved roughly  $\propto 1/\sqrt{N_{\rm slice} L^3}$,
where $L$ is lattice spatial dimension, and $N_{\rm slice}$ is the
number of time slices calculated light propagators.
In this work, we adopt lattice ensembles with relatively large $L$,
and sum $\pi\pi$ correlators over all time slices.
As a result, the relevant signals are indeed good enough.
Admittedly, most efficient way is used in Ref.~\cite{NPLQCD:2011htk}.

We compute two-point pion correlators with the zero and none-zero
momenta ($\mathbf{0}$ and $\mathbf{p}$) as well,
\begin{eqnarray}
\label{eq:pi_cor_PW_k000}
C_\pi({\mathbf 0}, t) &=& \frac{1}{T}\sum_{t_s=0}^{T-1}
\langle 0|\pi^\dag ({\mathbf 0}, t+t_s) W_\pi({\mathbf 0}, t_s) |0\rangle, \\
\label{eq:pi_cor_PW_k100}
C_\pi({\mathbf p}, t) &=& \frac{1}{T}\sum_{t_s=0}^{T-1}
\langle 0|\pi^\dag ({\mathbf p}, t+t_s) W_\pi({\mathbf p}, t_s) |0\rangle,
\end{eqnarray}
where $\pi$ and  $W_\pi$ are pion point-source and wall-source operator, respectively~\cite{Bernard:2001av,Aubin:2004wf},
and the summation over lattice space point in sink is not written.
Note that summations over all time slices $N_{\rm slice}$
ensure lattice-determined pion mass with desirable precision~\cite{Fu:2016itp}.

The pion mass $m_\pi$ and energy $E_\pi({\mathbf p})$ can be robustly obtained
at large $t$ from Eqs.~(\ref{eq:pi_cor_PW_k000})
and (\ref{eq:pi_cor_PW_k100}), respectively~\cite{Bazavov:2009bb},
\begin{eqnarray}
\label{eq:pi_fit_PW_k000}
\hspace{-0.6cm} C_\pi({\mathbf 0}, t) &=& A_\pi(\mathbf{0}) \left[e^{-m_\pi t}+e^{-m_\pi(T-t)}\right] +\cdots, \\
\label{eq:pi_fit_PW_k100}
\hspace{-0.6cm} C_\pi({\mathbf p}, t) &=& A_\pi(\mathbf{p})
\left[e^{-E_\pi({\mathbf p}) t}+e^{-E_\pi({\mathbf p})(T-t)}\right] + \cdots,
\end{eqnarray}
where the ellipses imply the oscillating parity partners,
and overlapping amplitudes $A_\pi(\mathbf{0})$ and $A_\pi(\mathbf{p})$
are later used to compute the wrap-around contributions~\cite{Gupta:1993rn,Umeda:2007hy}.

For the moving frame with only ground state, the energy $E_{\pi\pi}$  is extracted  from  $\pi\pi$ four-point function
which demonstrates as~\cite{Golterman:1985dz,DeTar:2014gla}
\begin{eqnarray}
\label{eq:E_pionpion}
\hspace{-1.0cm} C_{\pi\pi}(t)  &=&
Z_{\pi\pi}\cosh\left[E_{\pi\pi}\left(t - \frac{T}{2}\right)\right] \cr
&& + (-1)^t Z_{\pi\pi}^{\prime}\cosh \left[E_{\pi\pi}^{\prime} \left(t-\frac{T}{2}\right)\right] + \cdots.
\end{eqnarray}
for a large $t$ to suppress excited states,
the terms alternating in sign is due to a staggered scheme.
In our practice, the pollution by the ``wraparound'' effects~\cite{Gupta:1993rn,Umeda:2007hy,Feng:2009ij}
should be considered~\cite{Fu:2012gf,Fu:2013ffa}.
In practice, pion decay constants $f_\pi$ can be efficiently estimated by the approach in Ref.~\cite{Beane:2005rj},
which are listed in Table~\ref{tab:MILC_configs}.

As practiced in our previous works~\cite{Fu:2012tj,Fu:2011xw,Fu:2013ffa,Fu:2016itp},
for MF2 and MF4, the ground and first excited states can be separated
with variational method~\cite{Luscher:1990ck}
by calculating a $2 \times 2 $ correlation function matrix
$C(t)$ whose components can be estimated by Eq.~(\ref{eq:dcr}).
For this purpose, we construct a ratio of the correlation function matrices as
\begin{equation}
M(t,t_R) = C(t)  \, C^{-1}(t_R)  ,
\label{eq:M_def}
\end{equation}
with some reference time $t_R$~\cite{Luscher:1990ck}
to extract two lowest energy eigenvalues $\overline{E}_n$ ($n=1,2$),
which can be obtained by a fit to two eigenvalues
$\lambda_n (t,t_R)$ ($n=1,2$) of the correlation matrix $M(t,t_R)$~\cite{Gupta:1993rn,Golterman:1985dz,DeTar:2014gla,Fu:2012tj,Fu:2011xw,Fu:2013ffa,Fu:2016itp,Feng:2009ij,Umeda:2007hy},
\begin{eqnarray}
\label{Eq:asy}
\lambda_n (t, t_R) &=&
       A_n \cosh\left[-E_n\left(t-\frac{T}{2}\right)\right] \cr
       &&+
(-1)^t B_n \cosh\left[-E_n^{\prime}\left(t-\frac{T}{2}\right)\right] ,
\end{eqnarray}
for a large $t$ to suppress the excited states.

%
%
\section{FITTING ANALYSES}
\label{sec:pipiscattering}
\subsection{Lattice Phase Shift}
In order to intuitively display lattice measurements,
one usually computes the ratios~\cite{Kuramashi:1993ka,Fukugita:1994ve},
\begin{eqnarray}
\label{EQ:ratio}
R^X(t) = \frac{ C_{\pi\pi}^X(0,1,t,t+1) }{ C_\pi (0,t) C_\pi(1,t+1) }, \quad  X = D, C,\, {\rm and} \,\,C^W , \nonumber
\end{eqnarray}
where $C_\pi (0,t)$ and $C_\pi (1,t+1)$ are the pion correlators  with
an assigned momentum, and $C^W$ indicates the cross amplitude subtracted by wraparound  contamination.

\begin{figure}[!t]
\includegraphics[width=8.5cm,clip]{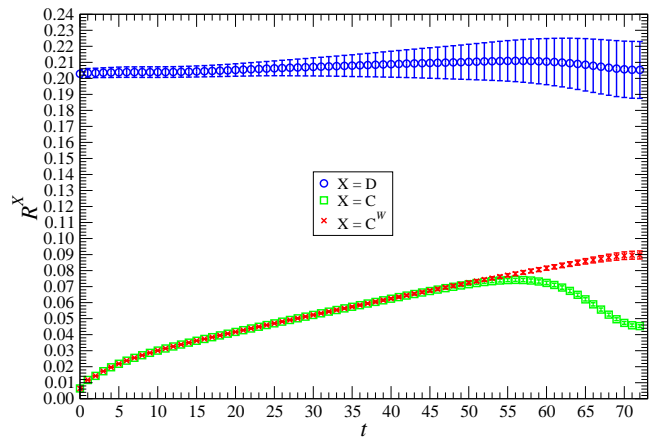}
\caption{\label{fig:rcd_r_000_I2}
(color online). Individual amplitude ratios $R^X(t)$ of $\pi\pi$ four-point functions computed
by the moving wall source technique at $\mathbf{P}=[0,0,0]$ for lattice ensemble $(0.0054,0.018)$:
the direct diagram (blue circle) displaced by $0.8$,
crossed diagram (green octagon), and crossed diagram (red cross) subtracted by wraparound contamination.
}
\end{figure}

Two contributions to an example $\pi\pi$ correlator for a lattice ensemble $(0.0054,0.018)$
at $\mathbf{P}=[0,0,0]$ as the functions of $t$ are
illustrated in Fig.~\ref{fig:rcd_r_000_I2},
which are shown as individual ratios $R^X, \, X=D, C,\, {\rm and} \,\,C^W $.
Note that some ratios for ensemble $(0.0031,0.031)$ are already displayed in Fig.~2
of the former work~\cite{Fu:2017apw}.
The ratio values of the direct amplitude $R^D$ are close to oneness,
hinting a quite weak interaction~\cite{Kuramashi:1993ka,Fukugita:1994ve,Fu:2012gf,Fu:2013ffa}.
As a matter of fact, $I=2$ $\pi\pi$  scattering is perturbative
at low momenta and small quark masses, as imposed by $\chi$PT.
As a result, two-pion energies deviate very slightly from the noninteracting energies~\cite{Kuramashi:1993ka,Fukugita:1994ve,Fu:2012gf,Fu:2013ffa}.

The systematically oscillating characteristics of the cross amplitude are difficult to be  viewed~\cite{Fu:2012gf,Fu:2013ffa},
which is a typical feature of the lattice staggered fermions~\cite{Golterman:1985dz}.
In this investigation, Prof. DeTar's strategy in Eq.~(\ref{eq:E_pionpion})
is normally manipulated to neatly remove this unwanted oscillating contribution~\cite{DeTar:2014gla}.

In the current study, lattice-measured quantities from two-point
correlations are adequately precise to enable us to nicely remove the wraparound pollution.
The detailed procedures for computing  the unwanted ``wraparound''  contamination in the
CoM frame and moving frames are elaborated in the former work~\cite{Fu:2016itp}.
From Fig.~\ref{fig:rcd_r_000_I2}, the contribution from
the finite-$T$ effects is obviously observed as $t$ approaches to $T/2=72$,
on the same time, it is nice to see that the ratios of  $C^W$ are almost in a straight line up to $t=72$.
Consequently, it is highly desirable for one to eliminate the ``wraparound''  contamination
with lattice-calculated wraparound contribution~\cite{Gupta:1993rn,Umeda:2007hy,Feng:2009ij} before fitting.

\begin{figure}[t!]
\includegraphics[width=8.5cm,clip]{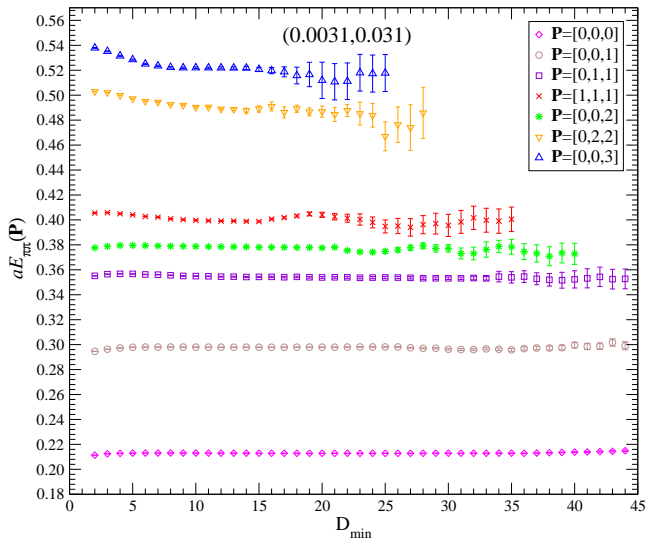}
\caption{\label{fig:eff_eng_I0} (color online).
Effective energy $E_{\pi\pi}$ plots for the lattice ensemble $(0.0031,0.031)$ as the functions of $\rm D_{min}$
for the $\pi\pi$ scattering in the $I=2$ channel in lattice units.
}
\end{figure}

A convincing technique to deal with lattice data is to
count on the ``effective energy'' plot~\cite{Bernard:2001av,Fu:2013ffa},
which is pretty analogous to the ``effective scattering length'' plot~\cite{Beane:2007xs}.
In practice, the $I=2$ $\pi\pi$ four-point correlators were fit by
modifying minimum fitting distances $\rm D_{min}$,
and setting maximum distance $\rm D_{max}$ either at $T/2$
or where the fractional statistical errors pass about $20\%$
for two sequential time slices~\cite{Bernard:2001av}.
An example ``effective energy'' plot for lattice $(0.0031,0.031)$
as the function of $\rm D_{min}$
is illustrated in Fig.~\ref{fig:eff_eng_I0}.
For $\mathbf{P}=[0,0,0]$, it is quite impressive that the plateau is
evidently watched from $\rm D_{min} = 5 \sim 40$.
%

The energies $a E_{\pi\pi}$ of $\pi\pi$ system
are secured from the ``effective energy'' plots,
and chosen by looking for a combination of a ``plateau'' in the energy
as the function of  $\rm D_{min}$, a good confidence level
and  $\rm D_{min}$ large enough to suppress the excited states~\cite{Beane:2007xs,Feng:2009ij}.

\begin{table*}[th!]	
\caption{\label{tab:pp_Io_kcotk}
Summaries of the lattice results for the fitted energies $E_{\pi\pi}$ of the $I=2$ $\pi\pi$ system.
For each total momentum $\mathbf{P}$, energy level, two single-pion are denoted by $({\mathbf p}_1, {\mathbf p}_2)$.
The fifth block shows the fitted energies $E_{\pi\pi}$ in lattice units.
Column six gives the fitting range,
and Column seven indicates the number of degrees of freedom (dof) for the fit.
The eight block is the center-of-mass scattering momentum $k^2$ in terms of $m_\pi^2$,
and Column nine gives the values of $k \cot\delta/m_\pi$, which are calculated
by L\"uscher formula~(\ref{eq:CMF}) or (\ref{eqn:RumGott}).
}
\begin{ruledtabular}
\begin{tabular}{lllllllll}
$\rm Ensemble$  & $\mathbf{P}$   & {\rm Level} & $({\mathbf p}_1, {\mathbf p}_2)$ & $a E$  &  {\rm Range} & $\chi^2/{\rm dof}$
& $k^2/m_\pi^2$   & $k\cot\delta/m_\pi $   \\
\hline
\multirow{9}*{$(0.005, 0.005)$}
&$[0,0,0]$ &$0$& $([0,0,0],[0,0,0])$ & $0.32377(12)$  & $22-32$ & $12.4/7$   & $0.01410(92)$  & $-7.09(46)$ \\
&$[0,0,1]$ &$0$& $([0,0,0],[0,0,1])$ & $0.41764(19)$  & $15-32$ & $26.7/14$  & $0.3144(16)$   & $-7.25(47)$ \\
&$[0,1,1]$ &$0$& $([0,0,0],[0,1,1])$ & $0.48456(37)$  & $14-32$ & $22.4/15$  & $0.5264(35)$   & $-5.70(74)$ \\
&          &$1$& $([0,1,0],[0,0,1])$ & $0.50982(19)$  & $10-32$ & $21.6/19$  & $0.7685(20)$   & $-7.73(55)$ \\
&$[1,1,1]$ &$0$& $([0,0,0],[1,1,1])$ & $0.53943(47)$  & $12-32$ & $24.5/17$  & $0.6960(50)$   & $-5.97(1.23)$ \\
&$[0,0,2]$ &$0$& $([0,0,1],[0,0,1])$~\footnote{We appreciate the work in Ref.~\cite{Bulava:2016mks}
for this kind of the energy level, which inspires us to use it, and  is very important for this work.
}
& $0.50848(20)$  & $13-32$ & $25.6/16$  & $0.00933(209)$ & $-6.72(1.48)$ \\
&          &$1$& $([0,0,0],[0,0,2])$ & $0.58677(56)$  & $10-25$ & $13.9/12$  & $0.8389(56)$   & $-8.98(2.43)$ \\
&$[0,2,2]$ &$0$& $([0,1,1],[0,1,1])$ & $0.64228(42)$  & $10-24$ & $12.4/11$  & $0.00717(529)$ & $-6.88(2.75)$ \\
&$[0,0,3]$ &$0$& $([0,0,1],[0,0,2])$ & $0.67945(65)$  & $11-22$ & $12.1/8$   & $0.1093(86)$   & $-5.95(1.88)$ \\
\hline
\multirow{8}*{$(0.0031, 0.031)$}
&$[0,0,0]$ &$0$ & $([0,0,0],[0,0,0])$ & $0.21285(9)$   & $28-48$ & $28.3/17$ & $0.02228(140)$   & $-8.22(51)$ \\
&$[0,0,1]$ &$0$ & $([0,0,0],[0,0,1])$ & $0.29814(18)$  & $25-44$ & $22.8/16$ & $0.4489(27)$     & $-6.64(38)$ \\
&$[0,1,1]$ &$0$ & $([0,0,0],[0,1,1])$ & $0.35418(59)$  & $17-48$ & $29.0/28$ & $0.7171(95)$     & $-6.18(1.20)$ \\
&$[1,1,1]$ &$0$ & $([0,0,0],[1,1,1])$ & $0.39873(46)$  & $15-48$ & $32.3/30$ & $0.9170(83)$     & $-9.68(2.56)$ \\
&$[0,0,2]$ &$0$ & $([0,0,1],[0,0,1])$ & $0.37877(15)$  & $11-48$ & $42.9/34$ & $0.01026(138)$   & $-9.64(1.09)$ \\
&$[0,2,2]$ &$0$ & $([0,1,1],[0,1,1])$ & $0.49025(45)$  & $11-32$ & $12.1/18$ & $0.00900(1008))$ & $-8.51(4.26)$ \\
&$[0,0,3]$ &$0$ & $([0,0,1],[0,0,2])$ & $0.52197(65)$  & $11-27$ & $23.6/13$ & $0.1369(154)$    & $-4.76(1.34)$ \\
\hline
\multirow{8}*{$(0.0031, 0.0031)$}
&$[0,0,0]$ &$0$ &$([0,0,0], [0,0,0])$ & $0.21285(9)$   & $26-48$ & $25.7/19$ & $0.02259(143)$   & $-8.118(473)$ \\
&$[0,0,1]$ &$0$ & $([0,0,0],[0,0,1])$ & $0.29812(18)$  & $25-46$ & $24.6/18$ & $0.4484(30)$     & $-6.79(48)$ \\
&$[0,1,1]$ &$0$ & $([0,0,0],[0,1,1])$ & $0.35421(65)$  & $17-50$ & $32.1/30$ & $0.7161(105)$    & $-6.25(1.76)$ \\
&$[1,1,1]$ &$0$ & $([0,0,0],[1,1,1])$ & $0.39890(50)$  & $14-48$ & $32.3/31$ & $0.9186(91)$     & $-9.04(3.76)$ \\
&$[0,0,2]$ &$0$ & $([0,0,1],[0,0,1])$ & $0.37877(16)$  & $11-48$ & $48.7/34$ & $0.01026(138)$   & $-8.91(1.03)$ \\
&$[0,2,2]$ &$0$ & $([0,1,1],[0,1,1])$ & $0.49021(44)$  & $11-33$ & $15.6/19$ & $0.00942(988))$  & $-8.14(3.94)$ \\
&$[0,0,3]$ &$0$ & $([0,0,1],[0,0,2])$ & $0.52198(65)$  & $11-27$ & $21.3/13$ & $0.1363(154)$    & $-4.83(1.38)$ \\
\hline
\multirow{11}*{$(0.0054, 0.018)$}
&$[0,0,0]$ &$0$ & $([0,0,0],[0,0,0])$ & $0.23163(14)$  & $38-50$ & $18.5/9$  & $0.02813(226)$ & $-3.21(24)$  \\
&$[0,0,1]$ &$0$ & $([0,0,0],[0,0,1])$ & $0.29353(30)$  & $29-53$ & $32.9/21$ & $0.3227(43)$   & $-2.64(18)$  \\
&$[0,1,1]$ &$0$ & $([0,0,0],[0,1,1])$ & $0.33496(50)$  & $25-54$ & $33.7/26$ & $0.4933(67)$   & $-3.02(79)$  \\
&          &$1$ & $([0,1,0],[0,0,1])$ & $0.35230(33)$  & $21-54$ & $36.4/30$ & $0.7217(86)$   & $-2.72(33)$  \\
&$[1,1,1]$ &$0$ & $([0,0,0],[1,1,1])$ & $0.37170(49)$  & $20-60$ & $42.8/37$ & $0.6625(72)$   & $-1.99(58)$  \\
&$[0,0,2]$ &$0$ & $([0,0,1],[0,0,1])$ & $0.34872(28)$  & $15-40$ & $28.5/22$ & $0.01688(424)$ & $-3.39(60)$  \\
&          &$1$ & $([0,0,0],[0,0,2])$ & $0.40302(88)$  & $18-35$ & $19.0/14$ & $0.7991(115)$  & $-2.81(93)$  \\
&$[0,2,2]$ &$0$ & $([0,1,1],[0,1,1])$ & $0.43575(56)$  & $20-48$ & $32.1/25$ & $0.0118(95)$   & $-3.81(1.54)$ \\
&$[0,0,3]$ &$0$ & $([0,0,1],[0,0,2])$ & $0.45952(74)$  & $19-36$ & $22.3/14$ & $0.1450(133)$  & $-1.89(51)$ \\
&$[2,2,2]$ &$0$ & $([1,1,1],[0,1,1])$ & $0.50836(63)$  & $16-46$ & $32.1/27$ & $0.0120(126)$  & $-3.2671.48)$ \\
&$[0,0,4]$ &$0$ & $([0,0,2],[0,0,2])$ & $0.57160(50)$  & $14-48$ & $41.2/31$ & $0.0074(111)$  & $-4.33(2.48)$ \\
\end{tabular}
\end{ruledtabular}
\end{table*}

The fitted values of the energies $E_{\pi\pi}$ in lattice units, fit range and fit quality ($\chi^2/{\rm dof}$)
are given in Table~\ref{tab:pp_Io_kcotk}.
Now it is straightforward to substitute fitted energies $E_{\pi\pi}$
into L\"uscher formula~(\ref{eq:CMF}) or (\ref{eqn:RumGott})
to secure the relevant values of the $k \cot \delta/m_\pi$.
All of these values are also summarized in Table~\ref{tab:pp_Io_kcotk},
where statistical errors of $k^2$ are calculated from
statistical errors of fit energies $E_{\pi\pi}$ and pion masses $m_\pi$ in lattice units.
Note that relevant scattering momenta $k$ are cited in units of
pion masses $m_\pi$ in order to explicate the remaining analysis
in such a manner which is independent of scale setting~\cite{NPLQCD:2011htk}.

We note that lattice-measured values of $k\cot\delta/m_\pi$ for each lattice ensemble
still have relatively large errors, which give rise to the relatively large statistical errors
for the relevant extracted quantities ($m_\pi a$, $m_\pi r$ and $P$).
The straightforward way to improve statistical error is to employ more gauge configurations
for each lattice ensemble~\cite{Lepage:1989hd,Fukugita:1994ve,Fu:2016itp},
which is definitely beyond the scope of this work
since it requires the huge of computer resources.

Moreover, according to the analytical arguments in Refs.~\cite{Lepage:1989hd,Fukugita:1994ve}
and a rule of thumb discussion in Ref.~\cite{Fu:2016itp},
most straightforward and efficient approach to improve relevant noise-to-signal ratios
is to employ anisotropic gauge configurations~\cite{NPLQCD:2011htk}
or lattice ensembles with large $L$,
and the signals are semi-empirically anticipated to be significantly improved~\cite{Fu:2016itp}.
We note that some other fitting strategies to improve the statistical errors
are discussed in Refs.~\cite{Lepage:2001ym,Hu:2017wli}.

\newpage
\subsection{The effective range approximation parameters}
\label{se:erap}
As it is explained latter, the effective range expansion ($\rm ERE$) is an expansion of real part
of inverse partial wave scattering amplitude in powers of
the magnitude of the CoM three-momentum of each pion,
namely,~\footnote{
Here we follow original derivations and notations
in Refs.~\cite{Gasser:1983yg,Bijnens:1995yn,Bijnens:1997vq,Colangelo:2001df}.
The definition of $P$ in Ref.~\cite{Fu:2017apw} is still adopted,
just because its NNLO expressions in $\chi$PT for $m_\pi r$ and $P$
are more easily to be derived, as shown later.
For a useful comprehension of effective range expansion, see Refs.~\cite{Adhikari:1983ii,Beane:2003da}.
}
\begin{equation}
\frac{k\,\cot{\delta}}{m_\pi} = \frac{1}{m_\pi a} + \frac{1}{2} m_\pi r\left(\frac{k^2}{m_\pi^2}\right)
                              + P  \left(\frac{k^2}{m_\pi^2}\right)^2 +\ldots,
\label{eq:effrange}
\end{equation}
where $m_\pi a$, $m_\pi r$ and $P$ are known as scattering length, effective range,
and shape parameter, respectively~\cite{NPLQCD:2011htk,Fu:2017apw}.
Note that the dimensionless quantity shape parameter $P$ is scaled with $(m_\pi r)^3$ in Ref.~\cite{NPLQCD:2011htk},
and there is no minus sign in first term in Eq.~(\ref{eq:effrange}),
which is consistent with notation in Eq.~(\ref{eq:threshexp}).
For simple notation, $m_\pi a \equiv m_\pi a_0^\mathrm{2}$, similar for $m_\pi r$ and $P$.

As pointed out in Ref.~\cite{NPLQCD:2011htk}, the effective range approximation is believed to be convergent
for energies below $t$-channel cut, which starts at ${k^2}={m_\pi^2}$.

\begin{figure}[t!]
\includegraphics[width=8.5cm,clip]{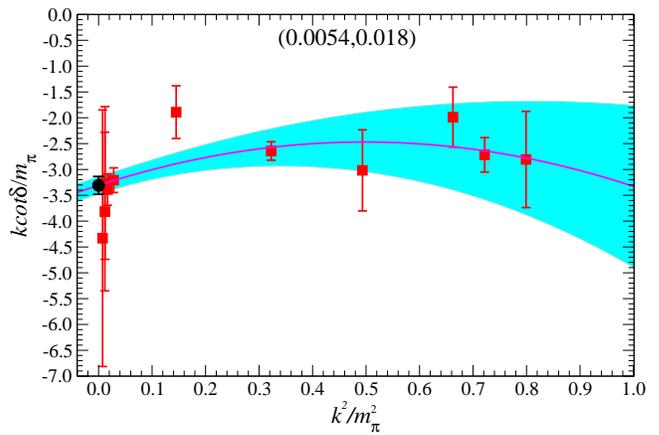}
\caption{
A three-parameter fit to lattice-determined values of $k\cot\delta/m_\pi$ over region $k^2/m_\pi^2<1.0$
for $(0.0054,0.018)$  ensemble.
The shaded cyan band corresponds to statistical error, and
solid magenta curve is the central values.
The black circle in this figure indicates the relevant fit value of $1/(m_\pi a)$.
}
\label{fig:pipiEREfitA}
\end{figure}

Due to the limited resources of computer,
only $7-11$ lattice data are at disposal for each lattice ensemble.~\footnote{
Most of the calculations for $(0.0031, 0.031)$ and $(0.0031, 0.0031)$ ensembles are actually already done
in our previous work~\cite{Fu:2017apw}, since we here can just pick up its $D$ and $C$ parts.
In spite of this fact, numerical computations of this study
were still unceasingly carried out more than six years.
}
In addition, lack of lattice ensembles with different $L$
for a given pion mass is an another important cause~\cite{NPLQCD:2011htk}.
Lattice-determined values of $k \cot{\delta}/m_\pi$
are summarized in Table~\ref{tab:pp_Io_kcotk},
where, in reality, only those  within $t$-channel cut ${k^2}={m_\pi^2}$  are tabulated.
An example of these values is exhibited in Fig.~\ref{fig:pipiEREfitA} for $(0.0031, 0.031)$ ensemble.

Moreover, we observe that the values of $k\cot\delta/m_\pi$ are
not roughly linear in $k^2$ during the region $k^2/m_\pi^2<1.0$,
which reflects the fact that the shape parameter $P$
indeed has a impact on the curvature~\cite{NPLQCD:2011htk}.
Actually, according to the quantitatively analytical discussions in Ref.~\cite{NPLQCD:2011htk}
and extended arguments in Sec.~\ref{sec:chiextrap},
the second term and third term in Eq.~(\ref{eq:effrange})
both contribute significantly for values of $k\cot\delta/m_\pi$~\cite{NPLQCD:2011htk}.

In region $k^2/m_\pi^2<1$, lattice measurements indicate that
curvatures have quadratic (and higher) dependence on $k^2$.
Hence, three leading effective range expansion parameters in Eq.~(\ref{eq:effrange})
are fit to lattice evaluations of $k\cot\delta/m_\pi$~\cite{NPLQCD:2011htk}.
Fitted values of $m_\pi a$, $m_\pi r$ and $P$ are given in the third Column of Table~\ref{tab:fitstoERT}.
An example fit of lattice computations is illustrated in Fig.~\ref{fig:pipiEREfitA} for $(0.0054, 0.018)$ lattice ensemble,
where the shaded cyan band corresponds to statistical error,
the solid magenta curve is the central values,
and the black circle manifests the relevant fit value of $1/(m_\pi a)$.

\begin{table}[t!]
\caption{Summaries of the effective range expansion parameters evaluated
from lattice determinations of $k\cot\delta/m_\pi$ for four lattice ensembles.
}
\label{tab:fitstoERT}
\begin{ruledtabular}
\begin{tabular}{l  c  l}
$\rm Ensemble$ & Quantity  &     Fit    \\
\hline
\multirow{4}*{$(0.0031, 0.031)$}
&$m_\pi a$          &  -0.1183(61)   \\
&$m_\pi r$          &  16.09(4.28) \\
&$P$                &  -8.15(3.50) \\
&$\chi^2/{\rm dof}$ &  6.14/4      \\
\hline
\multirow{4}*{$(0.0031, 0.0031)$}
&$m_\pi a$          & -0.1199(59)  \\
&$m_\pi r$          & 15.52(3.21) \\
&$P$                & -8.88(1.88) \\
&$\chi^2/{\rm dof}$ &  4.71/4     \\
\hline
\multirow{4}*{$(0.005, 0.005)$}
&$m_\pi a$          & -0.1396(75) \\
&$m_\pi r$          & 6.22(3.69)  \\
&$P$                & -4.51(2.34) \\
&$\chi^2/{\rm dof}$ &  5.48/6     \\
\hline
\multirow{4}*{$(0.0054, 0.018)$}
&$m_\pi a$          & -0.3023(158) \\
&$m_\pi r$          &  6.75(1.84)  \\
&$P$                & -3.40(1.26)  \\
&$\chi^2/{\rm dof}$ &  6.03/8      \\
\end{tabular}
\end{ruledtabular}
\end{table}

Admittedly, fitted values of effective range $m_\pi r$ and shape parameter $P$ listed in Table~\ref{tab:fitstoERT}
contain roughly reasonable statistical errors,
therefore, it is tolerable to use these data to perform the chiral extrapolation to the physical point.
For more sophisticated lattice study in the future,
it is highly desirable for one to produce more reliable data in the region $k^2/m_\pi^2<1.0$.
In what follows, $\chi$PT is exploited to predict the effective range expansion parameters at the physical point
with fit parameters listed in Table~\ref{tab:fitstoERT} as input.

%
%
%
\newpage
\section{Chiral  extrapolations}
\label{sec:chiextrap}
The low-energy theorems imposed by chiral symmetry reveal that
each scattering parameter can be associated to the corresponding LEC's
which appears in $\chi$PT~\cite{NPLQCD:2011htk}.
A state-of-the-art of this technique is demonstrated by NPLQCD
for isospin-$2$ $\pi\pi$ scattering at NLO in $\chi$PT~\cite{NPLQCD:2011htk}.
In this section, we will further extend NPLQCD's work in Ref.~\cite{NPLQCD:2011htk},
and provide its NNLO $\chi$PT expressions.

\subsection{Threshold parameters in $\chi$PT}
\label{sec:TPCHIPT}
In the elastic region, $I=2$  $\pi\pi$ $s$-wave scattering amplitude
$t(s)\equiv t_{\ell=0}^{I=2}(s)$ is associated with phase shift $\delta$~\cite{Gasser:1983yg,Bijnens:1997vq}
\begin{eqnarray}
t(s)= \Big(\frac{s}{s-4}\Big)^{\frac{1}{2}}\frac{1}{2i} \left\{e^{2i\delta(s)}-1\right\}\ .
\label{eq:unitary}
\end{eqnarray}
where Mandelstam variable $s=4(1+k^2/m_\pi^2)$ is denoted
in units of pion mass squared $m_\pi^2$~\cite{Gasser:1983yg,Bijnens:1997vq},
and $k$ is the magnitude of the CoM three-momentum of each pion.

In Appendix~\ref{app:ChPT NNLO}, the detailed procedures are provided to expand all terms of
the NNLO isospin-$2$ $\pi\pi$ scattering amplitude~\cite{Gasser:1983yg,Bijnens:1997vq,Colangelo:2001df}
in powers of $k^2$(consult Appendix~\ref{app:ChPT NNLO} for more details),
consequently, Taylor series of $\pi\pi$ scattering amplitude in $k^2$ has a pretty concise form.
It is obvious that nine independent LEC's $C_i(i=1-9)$~\cite{Gasser:1983yg,Bijnens:1997vq}
are needed to describe NNLO scattering amplitude $t(k)$ as
\begin{eqnarray}
\label{eq:pipi_nnlo}
\hspace{-0.6cm}t_0^2(k)&=&
-\frac{m_\pi^2}{8\pi f_\pi^2} + \frac{m_\pi^4}{f_\pi^4}C_1
+\frac{m_\pi^6}{f_\pi^6}C_4 \cr
&&{}\cr
&&+
\frac{k^2}{f_\pi^2}\bigg[-\frac{1}{4\pi} + \frac{m_\pi^2}{f_\pi^2} C_2
+\frac{m_\pi^4}{f_\pi^4}C_5\bigg] \cr
&&{}\cr
&&+\frac{k^4}{f_\pi^4}\left[ C_3 + \frac{m_\pi^2}{f_\pi^2}C_6\right] \cr
&&{}\cr
&&+
\frac{m_\pi^6}{f_\pi^6}\left[ C_7 + C_8\frac{k^2}{m_\pi^2} + C_9\frac{k^4}{m_\pi^4} \right]
\ln\left(\frac{m_\pi^2}{f_\pi^2}\right) \cr
&&{}\cr
&&-
\frac{m_\pi^4}{4\pi^3 f_\pi^4}\left[ \frac{3}{32} + \frac{5}{12} \frac{k^2}{m_\pi^2}
+\frac{5}{9} \frac{k^4}{m_\pi^4} \right]\ln\hspace{-0.05cm}
\left(\frac{m_\pi^2}{f_\pi^2}\right) \cr
&&{}\cr
&&+
\frac{m_\pi^6}{64 \pi^5 f_\pi^6}\hspace{-0.05cm}\left[ \frac{31}{24}
\hspace{-0.05cm}+\hspace{-0.05cm} \frac{169}{144} \frac{k^2}{m_\pi^2}
\hspace{-0.05cm}-\hspace{-0.05cm}\frac{35}{27}\frac{k^4}{m_\pi^4} \right]\hspace{-0.05cm}
\ln^2\hspace{-0.1cm}\left(\frac{m_\pi^2}{f_\pi^2}\right)\hspace{-0.05cm},
\end{eqnarray}
where $C_i(i=1-9)$ can be conveniently expressed in terms of low-energy constants  $\tilde{\ell}_i(\mu=f_{\pi,\mathrm{phy}})$~\cite{Gasser:1983yg,Bijnens:1997vq,Colangelo:2001df}
\begin{eqnarray}
\label{eq:c12345_r}
C_1 &=&
\frac{1}{128\pi^3}\left[\frac{8}{3}\tilde{\ell}_{1} +\frac{16}{3}\tilde{\ell}_{2} - \tilde{\ell}_{3} - 4\tilde{\ell}_{4} + 1 \right]\cr
%
&&{}\cr
C_2 &=&
\frac{1}{64\pi^3}\left[\frac{8}{3}\tilde{\ell}_1 + 8\tilde{\ell}_2-4\tilde{\ell}_4 - \frac{5}{6}\right]\cr
%
&&{}\cr
C_3 &=&
\frac{1}{36\pi^3}\left[\tilde{\ell}_1 + 4\tilde{\ell}_2 -\frac{193}{40}\right]\cr
&&{}\cr
C_4 &=&
\frac{1}{512\pi^5}\bigg[\frac{16}{3}\tilde{\ell}_{1}\tilde{\ell}_{4} +
\frac{32}{3}\tilde{\ell}_{2}\tilde{\ell}_{4}-3\tilde{\ell}_{3}\tilde{\ell}_{4} -5\tilde{\ell}_{4}^2
-\frac{1}{2}\tilde{\ell}_3^{2}- \cr
&&\frac{4}{3}\tilde{\ell}_{1} - \frac{16}{3}\tilde{\ell}_{2}-\frac{7}{4}\tilde{\ell}_{3} - \tilde{\ell}_{4}
-{\frac{163}{16}}+\frac{22}{9}\pi^2 + \tilde{r}_1 + 16 \tilde{r}_4 \bigg]\cr
%
&&{}\cr
C_5 &=&
\frac{1}{512\pi^5}\bigg[ \frac{32}{3}\tilde{\ell}_1\tilde{\ell}_4
+ 32\tilde{\ell}_2\tilde{\ell}_4 - 2\tilde{\ell}_3 \tilde{\ell}_4 - 10\tilde{\ell}_4^{2} \cr
&&+\frac{92}{27}\tilde{\ell}_1 - \frac{340}{27}\tilde{\ell}_2
  - \frac{43}{6}\tilde{\ell}_3 -\frac{50}{3}\tilde{\ell}_4 + \frac{317}{27}\pi^2
  -\frac{13481}{324} \cr
&&- 2\tilde{r}_2 + 48\tilde{r}_4 - 32\tilde{r}_6 \bigg]\cr
%
&&{}\cr
C_6 &=&
\frac{1}{96\pi^5}\bigg[\tilde{r}_3 + 7\tilde{r}_4 -16\tilde{r}_6
+\frac{4}{3}\tilde{\ell}_1\tilde{\ell}_4+ \frac{16}{3}\tilde{\ell}_2\tilde{\ell}_4+\frac{21}{5}\tilde{\ell}_1\cr
&&+\frac{116}{45}\tilde{\ell}_2-\frac{89}{120}\tilde{\ell}_3-\frac{293}{30}\tilde{\ell}_4
+\frac{4453}{1440}\pi^2-\frac{5635}{1296} \bigg]\cr
%
&&{}\cr
C_7 &=&
\frac{1}{512\pi^5}\left[-\frac{4}{3}\tilde{\ell}_{1}-8\tilde{\ell}_{2} {+} \tilde{\ell}_{3}-2\tilde{\ell}_{4}
+\frac{155}{12}\right]\cr
%
&&{}\cr
C_8 &=&
\frac{1}{512\pi^5}\left[\frac{80}{9}\tilde{\ell}_1-4\tilde{\ell}_2
+5\tilde{\ell}_3-\frac{56}{3}\tilde{\ell}_4+\frac{3041}{54}\right]\cr
%
&&{}\cr
C_9 &=&
\frac{1}{96\pi^5}\left[\frac{41}{9}\tilde{\ell}_1 + 6\tilde{\ell}_2 -\frac{20}{3}\tilde{\ell}_4
+\frac{9349}{1080}\right],
\end{eqnarray}
where the coupling constants $\tilde{\ell}_i$ and $\tilde{r}_i$ are in $\chi$PT of the chiral order $4$ and $6$
renormalized at the physical pion decay constant ($f_{\pi,\mathrm{phy}}$) quoted by PDG~\cite{Workman:2022ynf}.

Note that all low-energy constants are at the scale of
physical pion decay constant (see Appendix~\ref{app:C NNLO}
for more details) in favor of NPLQCD's notations
and are independent of quark mass, such that one can take them
as fitting parameters in the chiral extrapolations.
It is worth emphasizing that 
if dropping all $m_\pi^6/f_\pi^6$ terms in Eq.~(\ref{eq:pipi_nnlo}),
it neatly reduces to its NLO expression in Ref.~\cite{NPLQCD:2011htk}.

Near threshold behavior of partial wave amplitude $t(k)$
can be written as a power-series expansion in the CoM three-momentum~\cite{Gasser:1983yg,Bijnens:1997vq}
\begin{eqnarray}
\mbox{Re}\; t(k)\ =\ m_\pi a +k^2 b + k^4 c + {\cal O}(k^6),
\label{eq:threshexp}
\end{eqnarray}
where $a$ is referred to as the scattering length, $b$ and $c$
are called as slope parameters, respectively.
Note that there is no minus sign before $m_\pi a$ in Eq.~(\ref{eq:threshexp}) as it in Ref.~\cite{NPLQCD:2011htk}, which just in order to follow the notations in Ref.~\cite{Bijnens:1997vq},
and  be consistent with definition of the effective range approximation in Eq.~(\ref{eq:effrange}).

Matching the threshold expansion in Eq.~(\ref{eq:threshexp}) to
ERE in Eq.~(\ref{eq:effrange}), the effective range $r$ and
shape parameter $P$ can be neatly described
just in terms of three threshold parameters~\cite{Bijnens:1997vq}:
\begin{eqnarray}
m_\pi r &=& \frac{1}{m_\pi a}-\frac{2 m^2_\pi b}{(m_\pi a)^2} - 2 m_\pi a, \label{eq:m_pi_r} \\
P &=& \frac{1}{2}m_\pi a - \left(m_\pi a\right)^3 -  m_\pi^2 b  - \frac{1}{8 m_\pi a} +\frac{\left(m_\pi^2 b\right)^2}{(m_\pi a)^3}\cr
&&-\frac{m_\pi^2 b}{2(m_\pi a)^2}-\frac{m_\pi^4 c}{(m_\pi a)^2},
\label{eq:m_pi_P}
\end{eqnarray}
where the compact form of $P$ in Refs.~\cite{NPLQCD:2011htk,Fu:2017apw} is recast into seven separate terms,
which are actually not all relevant to the final NLO representations in $\chi$PT~\cite{NPLQCD:2011htk}.
As a matter of fact, the NNLO expressions does not include as well all components.
Note that only $P$ depends on slope parameter $c$.

To simplify the notation, it is convenient to follow the notation
in Ref.~\cite{NPLQCD:2011htk} to denote $z\equiv m_\pi^2/f_\pi^2$.
From Taylor series of scattering amplitude in powers of $k^2$ in Eq.~(\ref{eq:pipi_nnlo}),
it is straightforward to obtain NNLO $\chi$PT expressions for threshold parameters:
\begin{eqnarray}
m_{\pi} a &=& -\frac{z}{8\pi} + z^2 C_1 - \frac{3}{128\pi^3} z^2 \ln z \cr
&&+ z^3 C_4 + z^3\ln z C_7 + \frac{31}{3072\pi^5}z^3(\ln z)^2, \label{eq:Thresholds_a} \\
m_{\pi}^2b&=&-\frac{z}{4\pi} + z^2C_2 -\frac{5}{48\pi^3}z^2\ln z \cr
&&+ z^3 C_5 + C_8 z^3\ln z + \frac{169}{9216\pi^5}z^3(\ln z)^2, \label{eq:Thresholds_b}  \\
m_{\pi}^4 c &=& z^2C_3 -\frac{5}{36\pi^3}z^2 \ln z \cr
&&+ z^3 C_6 + C_9 z^3\ln z -\frac{35}{1728\pi^5}z^3(\ln z)^2.
\label{eq:Thresholds_c}
\end{eqnarray}

As it demonstrated in Appendix~\ref{app:C NNLO},
Equation~(\ref{eq:Thresholds_a}) and (\ref{eq:Thresholds_b}) are neat reproductions
for the scattering length $a$ and slope parameter $b$ in Ref.~\cite{Bijnens:1997vq}.
It should be worthwhile to stress that
there is no explicit expression of the slope parameter $c$ in Ref.~\cite{Bijnens:1997vq}.
Admittedly, its NLO parts of $c$ are already indicated in Refs.~\cite{NPLQCD:2011htk,RBC:2021acc},
and it is very helpful to use them as a double-check
to the derivations of the NNLO components.

From Eq.~(\ref{eq:m_pi_r}), it is apparent that the effective range $m_\pi r$
comprises three components, and the third term dominates its behaviors for the large $z$-values,
which actually does not appear in the NLO expression~\cite{NPLQCD:2011htk}.
Since its NNLO representation includes more high order terms, it is natural to contain more
information from all three parts.

Analogously, from Eq.~(\ref{eq:m_pi_P}), the shape parameter $P$ involves seven terms,
first three components dominates its behaviors for the large $z$-values,
which also do not emerge in NLO expression~\cite{NPLQCD:2011htk},
and partially enter into the NNLO espression.
As a matter of fact,  any information of the second term in Eq.~(\ref{eq:m_pi_P}) 
does not enter the NNLO expression either.

Substituting Eqs.~(\ref{eq:Thresholds_a})-(\ref{eq:Thresholds_c})] into
Eqs.~(\ref{eq:m_pi_r})-(\ref{eq:m_pi_P}), after some strenuous algebraic manipulations,
it is relatively straightforward to achieve NNLO $\chi$PT descriptions for the effective range approximation parameters:
\begin{eqnarray}
m_\pi r \hspace{-0.05cm}&=&\hspace{-0.05cm} \frac{24\pi}{z} + C_{10} + \frac{17}{6\pi}\ln z \cr
&&+z C_{11} + z\ln z C_{12} + \frac{77}{288\pi^3}z(\ln z)^2
\label{eq:mpr_NNLO} \\
P &=& -\frac{23\pi}{z}+ C_{13} + \frac{53}{144\pi}\ln z \cr
&&+ z C_{14} +  z\ln z C_{15} -\frac{13711}{6912\pi^3} z\left(\ln z\right)^2
\label{eq:mP_NNLO}
\end{eqnarray}
where the constants $C_i(i=10-15)$  are solely denoted
in terms of the constants $C_i(i=1-9)$ via
\begin{eqnarray}
C_{10} &=& 64\pi^2\left(7C_1 - 2C_2 \right)  \cr
&&{}\cr
C_{11} &=& \frac{1}{4\pi} + 64\pi^2\left(7C_4 -2C_5- 32\pi C_1C_2+ 88\pi C_1^2\right)  \cr
&&{}\cr
C_{12} &=& 48C_2 - \frac{152}{3}C_1+448\pi^2C_7-128\pi^2C_8 \cr
&&{}\cr
C_{13} &=& 8\pi^2\left(28C_2-79C_1-8C_3 \right)\label{eq:extrapforms_C12}  \cr
&&{}\cr
C_{14} &=& \frac{3}{16\pi}+8\pi^2\left(-79C_4 + 28C_5 -8C_6 \right)\cr
&&{}\cr
&&+64\pi^3\left(-167 C_1^2 + 88C_1C_2 - 8 C_2^2 -16 C_1C_3\right)   \cr
&&{}\cr
C_{15} &=& \frac{509}{9}C_1 - \frac{76}{3} C_2 + 24C_3 \cr
&&{}\cr
&&- 632\pi^2 C_7 + 224\pi^2C_8 - 64\pi^2C_9.
\label{eq:newConst1015}
\end{eqnarray}

Note that first three terms in Eqs.~(\ref{eq:mpr_NNLO}) and~(\ref{eq:mP_NNLO})
are related to relevant NLO $\chi$PT expressions in Ref.~\cite{NPLQCD:2011htk},
and the constants $C_i(i=1,10,13)$ are three constants denoted at NLO in $\chi$PT~\cite{NPLQCD:2011htk}.
The rest parts in these equations are new features from NNLO in $\chi$PT.

Combining Eq.~(\ref{eq:mpr_NNLO}) and ~(\ref{eq:mP_NNLO}), it is easy to  testify that
\begin{equation}
\frac{P}{(m_\pi r)^{3}} = -\frac{23}{13824\pi^2}z^2 + z^3 C_{13}^\prime + \frac{613}{995328\pi^4}z^3\ln z + \cdots
\nonumber
\end{equation}
where $C_{13}^\prime \equiv \frac{1}{864\pi}\left(41C_1 -9C_2 -4C_3\right)$,
and the ellipsis indicates high order contributions.
With definitions in Eq.~(\ref{eq:c12345_r}),~\footnote{
The $\tilde{\ell}_i$ are connected with $\ell_1^r$ by equality
$
\ell_i^r = \frac{1}{32\pi^2} \gamma_i\tilde{\ell}_i ,
$
where $\gamma_1=\frac{1}{3}$, $\gamma_2=\frac{2}{3}$, $\gamma_3=-\frac{1}{2}$, and $\gamma_4=2$~\cite{Bijnens:1997vq,Colangelo:2001df}.
}
it is pretty nice and pleasure to show that
\begin{eqnarray}
C_{13}^\prime &=&
 \frac{1}{5184\pi^2}\left[ 212\ell_1^r + 40\ell_2^r
 + 123\ell_3^r \hspace{-0.05cm}-\hspace{-0.05cm}69\ell_4^r\right] + \frac{701}{622080\pi^4}, \cr
C_{10}       &=&
32\pi\left(12\ell_1^r + 4\ell_2^r + 7\ell_3^r -3\ell_4^r \right) + \frac{31}{6\pi},
\end{eqnarray}
which is neatly consistent with corresponding NPLQCD's definitions
of the constants $C_4$ and $C_2$, respectively~\cite{NPLQCD:2011htk}.
Moreover, since there is no minus sign in the first term in Eq.~(\ref{eq:effrange}),
when comparing with relevant $C_1$ in Ref.~\cite{NPLQCD:2011htk},
its counterpart should be multiplied by $(-1)$.
So far, we demonstrate that our expressions for $m_\pi r$ and $P$
are equivalent with their NPLQCD's at NLO in $\chi$PT.

In  principle, at least four lattice ensembles are required to determine
the nine independent low-energy constants.
In practice, if the enough lattice-measured effective range expansion parameters ($m_\pi a$, $m_\pi r$, and
$P$) are at hand, one can use Equations~(\ref{eq:m_pi_r}), (\ref{eq:m_pi_P}), and~(\ref{eq:Thresholds_a})
to fit the relevant lattice data to get nine constants: $C_1$, $C_4$, $C_7$, $C_{10}$,
$C_{11}$, $C_{12}$, $C_{13}$, $C_{14}$ and $C_{15}$, which then can be employed to acquire
the slope parameters $b$ and $c$.~\footnote{
This option is a natural aftermath of NPLQCD's work in Ref.~\cite{NPLQCD:2011htk},
which clearly have more actual physical meanings and is more convenient to fit relevant lattice data.
In practice, one can choose any suitable nine independent low-energy constants to fit their lattice data.
Moreover, $C_{10}$, $C_{11}$ and $C_{12}$ are associated to the effective range $r$,
and $C_{13}$, $C_{14}$ and $C_{15}$ shape the parameter $P$.
}
In the rest of the analysis, as it done in Ref.~\cite{NPLQCD:2011htk},
we refer to these nine constants as the independent constants instead of the constants
$C_i (i=1-9)$, which can be derived from them, namely,
from Eq.~(\ref{eq:newConst1015}), we get
\begin{eqnarray}
C_{2} &=&
\frac{7}{2}C_1- \frac{1}{128\pi^2} C_{10}   \cr
&&{}\cr
C_{3}&=&
\frac{19}{8} C_1 - \frac{7}{256\pi^2} C_{10}- \frac{1}{64\pi^2} C_{13} \cr
&&{}\cr
C_{5}&=&
\frac{1}{512\pi^3} + \frac{7}{2}C_4- \frac{1}{128\pi^2} C_{11} \cr
&&{}\cr
C_{6}&=&
\frac{19}{8} C_4 - \frac{1}{64\pi^2} C_{14} + \frac{31}{1024\pi^3} - \frac{7}{256\pi^2} C_{11} \cr
&&+43\pi C_1^2 + \frac{3}{16\pi}C_1C_{10}  -  \frac{1}{2048\pi^3} C_{10}^2
+\frac{1}{4\pi}C_1C_{13} \cr
&&{}\cr
C_{8}&=& \frac{11}{12\pi^2} C_1 + \frac{7}{2}C_7 - \frac{3}{1024\pi^4} C_{10}- \frac{1}{128\pi^2} C_{12} \cr
&&{}\cr
C_{9}&=& \frac{259}{72\pi^2} C_1 + \frac{19}{8} C_7 -\frac{11}{1536\pi^4} C_{10}
- \frac{7}{256\pi^2} C_{12} \cr
&&- \frac{3}{512\pi^2} C_{13} - \frac{1}{64\pi^2} C_{15}
\label{eq:newConst258369}
\end{eqnarray}
which is partially compatible with Eq.~(10) in Ref.~\cite{NPLQCD:2011htk}.
Meanwhile, using Eqs. (\ref{eq:c12345_r}) and (\ref{eq:newConst1015}), we can also recast
the $C_i (i=10-15)$ in terms of the scale-independent dimensionless couplings
\begin{eqnarray}
C_{10}&=& \frac{1}{6\pi}\left[  24 \tilde{\ell}_{1} + 16\tilde{\ell}_{2}
-21\tilde{\ell}_{3} -36 \tilde{\ell}_{4} + 31 \right] \cr
&&{}\cr
C_{11}&=& \frac{1}{8\pi^3}\bigg[\frac{15427}{1296} -\frac{118}{27}\pi^2 +16\tilde{\ell}_{1}\tilde{\ell}_{4}
+\frac{32}{3}\tilde{\ell}_{2}\tilde{\ell}_{4}-17\tilde{\ell}_{3}\tilde{\ell}_{4} \cr &&-15\tilde{\ell}_{4}^2-\frac{7}{2}\tilde{\ell}_3^{\ 2} - \frac{436}{27}\tilde{\ell}_{1}
- \frac{328}{27}\tilde{\ell}_{2}+\frac{25}{12}\tilde{\ell}_{3} +\frac{79}{3}\tilde{\ell}_4  \cr
&&+ 7\tilde{r}_1 + 4\tilde{r}_2 + 16 \tilde{r}_4 +64\tilde{r}_6\bigg]\cr
&&{}\cr
C_{12}&=& \frac{1}{9\pi^3}\left[ - 22 \tilde{\ell}_1 -19 \tilde{\ell}_2 + \frac{3}{16} \tilde{\ell}_3 + \frac{27}{2} \tilde{\ell}_4
- \frac{3281}{96}\right]\cr
&&{}\cr
C_{13}&=&
\frac{1}{720\pi}\left[521-4040\tilde{\ell}_{1} -3920\tilde{\ell}_{2}+3555\tilde{\ell}_{3} + 4140\tilde{\ell}_{4}\right]  \cr
&&{}\cr
C_{14}&=& \frac{1}{\pi^3}\bigg[  \frac{353}{1440}\pi^2- \frac{3694529}{1244160}-\frac{133}{108}\tilde{\ell}_{1}^2 - \frac{53}{27}\tilde{\ell}_{2}^2
-\frac{9}{256}\tilde{\ell}_{1}^3  \cr
&&{}\cr
&&+ \frac{23}{64}\tilde{\ell}_{4}^2 - 3\tilde{\ell}_{1}\tilde{\ell}_{2}
+\frac{269}{144}\tilde{\ell}_{1}\tilde{\ell}_{3}
+\frac{169}{72}\tilde{\ell}_{2}\tilde{\ell}_{3}
+\frac{23}{64}\tilde{\ell}_{3}\tilde{\ell}_{4} \cr
&&{}\cr
&&+\frac{16}{45}\tilde{\ell}_1 + \frac{289}{216}\tilde{\ell}_2 + \frac{3739}{11520}\tilde{\ell}_3
+ \frac{53}{576}\tilde{\ell}_4 \cr
&&{}\cr
&&-\frac{79}{64}  \tilde{r}_1 -\frac{7}{8}  \tilde{r}_2-\frac{2}{3}  \tilde{r}_3
-\frac{41}{12}  \tilde{r}_4 -\frac{10}{3}  \tilde{r}_6 \bigg] \cr
&&{}\cr
C_{15}&=&\frac{1}{12\pi^3}\hspace{-0.1cm}\left[ \frac{355}{9}\tilde{\ell}_1 + \frac{646}{9}\tilde{\ell}_2  + \frac{589}{96}\tilde{\ell}_3
- \frac{207}{12}\tilde{\ell}_4 +  \frac{14515}{1728} \right].\nonumber
\end{eqnarray}
It is interesting and important to note that all constants  $C_1, C_4, C_7, C_i (i=10-15)$
are solely dependent on nine coupling constants: $\tilde{\ell}_1,\tilde{\ell}_2,\tilde{\ell}_3,\tilde{\ell}_4,
\tilde{r}_1,\tilde{r}_2,\tilde{r}_3,\tilde{r}_4,\tilde{r}_6$~\cite{Gasser:1983yg,Bijnens:1997vq}.

Note that~\cite{Bijnens:1997vq,Colangelo:2001df}
$$
\tilde{\ell}_i = \bar{\ell}_i + \ln \frac{m_\pi^2}{\mu^2}, 
$$
where $\bar{\ell}_i$ are scale-independent dimensionless couplings.
Hence, $\tilde{\ell}_i$ can be estimated at any scale $\mu$.
In the present sudy, we quote published data of $\bar{\ell}_i$~\cite{Colangelo:2001df,Bijnens:2014lea},
\begin{eqnarray}
\bar{\ell}_1 &=& -0.4\pm0.6, \quad \bar{\ell}_2 = 4.3 \pm 0.1, \cr
\bar{\ell}_3 &=& 3.0\pm 0.8, \qquad  \bar{\ell}_4 = 4.3 \pm 0.3.
\nonumber
\end{eqnarray}

\begin{figure}[t]
\includegraphics[width=8.5cm,clip]{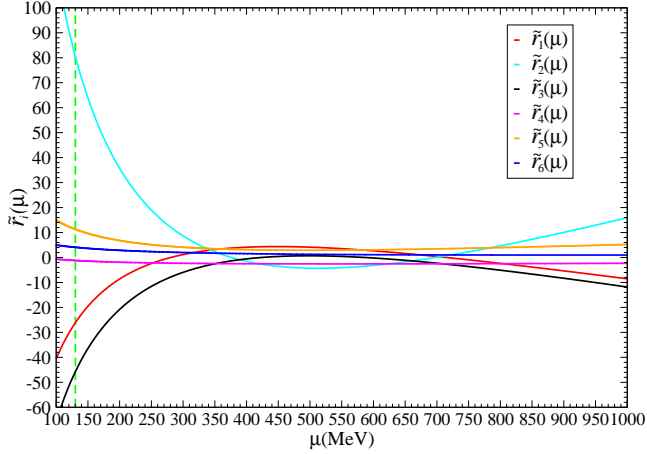}
\caption{
The scale dependence of $\tilde{r}_i(\mu)(i=1-6)$.
The dashed green line denotes the physical pion decay constant $f_{\pi,\rm phy}$.
}
\label{fig:cgl_NNLO_r16}
\end{figure}

Moreover, coupling constants $\tilde{r}_i$
are scale dependent, and complicated since it is quadratic
in $\ln\mu$~\cite{Bijnens:1997vq,Colangelo:2001df,Bijnens:2014lea}.
Its analytic expressions are implied
in Refs.~\cite{Bijnens:1997vq,Colangelo:2001df,Bijnens:2014lea},
in Appendix~\ref{app:R NNLO}, we explicitly offer corresponding scale dependent
quantities for convenience's sake. 
In principle,  with the assistance of the analytical expression in Eq.~(\ref{app_ChPT_R}),
one can get its value at any scale $\mu$,
where the integration constants can be fixed with the obtained $\tilde{r}_i(i=1-6)$ values
at a given scale: e.g., $\mu=M_\rho=0.77~{\rm GeV}$.
We instead quote the obtained values~\cite{Colangelo:2001df}.
\begin{eqnarray}
\tilde{r}_1(\rho)  &=& -1.5\,, \qquad \tilde{r}_2(\rho) = 3.2\,,  \cr
\tilde{r}_3(\rho)  &=& -4.2\,, \qquad \tilde{r}_4(\rho) = -2.5\,,  \cr
\tilde{r}_5(\rho)  &=& -3.8\,,  \qquad \tilde{r}_6(\rho) = 1.0\,.
\nonumber
\end{eqnarray}
The scale dependence of coupling constant $\tilde{r}_i(i=1-6)$
is displayed in Fig.~\ref{fig:cgl_NNLO_r16}.
Then the values obtained at the physical pion decay constant $f_{\pi,\rm phy}$~\cite{Workman:2022ynf} give
\begin{eqnarray}
\tilde{r}_1(f_{\pi,\rm phy})  &=& -27.2\,, \qquad \tilde{r}_2(f_{\pi,\rm phy}) = 81.8\,,  \cr
\tilde{r}_3(f_{\pi,\rm phy})  &=& -45.7\,, \qquad \tilde{r}_4(f_{\pi,\rm phy})= -1.2\,,  \cr
\tilde{r}_5(f_{\pi,\rm phy})  &=&  11.4\,, \,\,\,\,\,\qquad \tilde{r}_6(f_{\pi,\rm phy}) = 4.1\,.
\nonumber
\end{eqnarray}
The values of $C_i(i=1-15)$ are estimated at $f_{\pi,\rm phy}$  as
\begin{eqnarray}
C_1 &=& 0.000777(559)\,,   \quad\quad \,C_2 = 0.00817(109)\,,\cr
C_3 &=& 0.01135(64)\,,     \quad\quad\quad \,C_4 = -0.000036(173))\,,  \cr
C_5 &=& -0.000224(386)\,,  \quad\,\, \,C_6 = -0.000590(423)\,,\cr
C_7 &=& -0.000179(9)\,,    \quad\quad\,\,\,C_8 = -0.000398(50)\,,\cr
C_9 &=& 0.000153(120)\,,   \quad\quad C_{10} = -6.89(1.31), \cr
C_{11} &=&  0.203(361)\,,  \quad\quad\quad \,\,\, C_{12} = 0.0658(499)\,,\cr
C_{13} &=&  6.06(1.75)\,,  \quad\quad\quad\,\,\,\, C_{14} = -0.096(398)\,, \cr
C_{15} &=&  0.247(69)\,.
\label{eq:C123_CGL}
\end{eqnarray}

With the evaluated values in Eq.~(\ref{eq:C123_CGL}),
and using equations~(\ref{eq:m_pi_r}) and (\ref{eq:m_pi_P}),
the ratio of the effective range $m_\pi r$ to
shape parameter $P$ at physical point is computed as $-0.99(29)$.
Note that NPLQCD's lattice data in Ref.~\cite{NPLQCD:2011htk} indicates this ratio
is $-0.95(11)$.

\begin{figure}[t!]
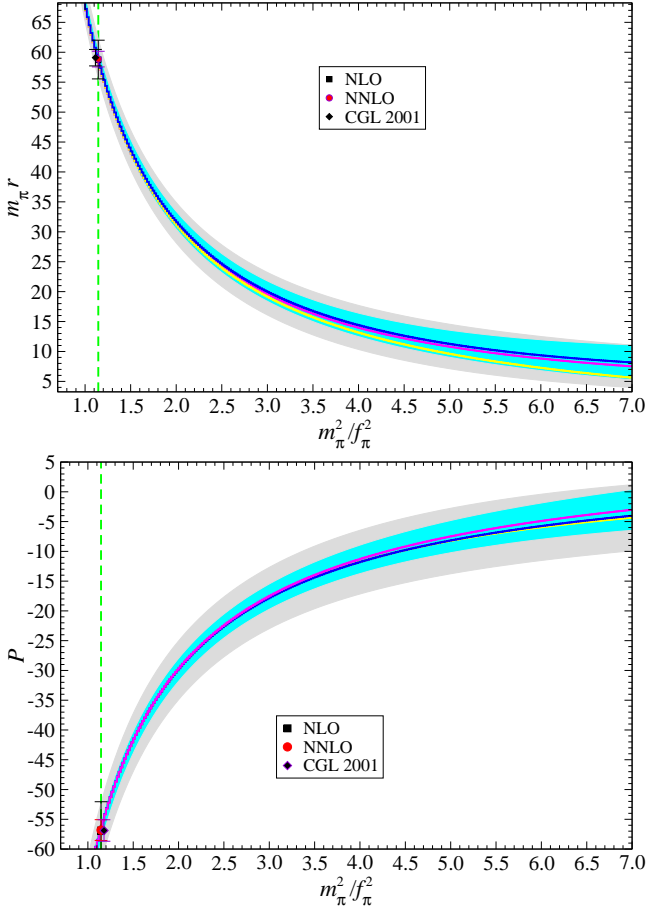

\includegraphics[width=8.5cm,clip]{NNLO_mr_fP.eps}
\includegraphics[width=8.5cm,clip]{fP_NNLO_P.eps}
\caption{
The dashed green lines denote the physical line,
and the Roy equation predictions~\protect\cite{Colangelo:2001df}
are the black circles on the physical line
by using Eq.~(15) in Ref.~\cite{NPLQCD:2011htk} with the relevant input data from Ref.~\cite{Colangelo:2001df}.
For comparison, the grey bands denotes the relevant results are directly evaluated
using Eq.~(\ref{eq:m_pi_r}) or (\ref{eq:m_pi_P}) as the references (CGL 2001).
The cyan bands shows the results are calculated at NNLO
using Eqs.~(\ref{eq:mpr_NNLO}) and (\ref{eq:mP_NNLO}).
The solid yellow, blue, and magenta curves are the central values for NLO, NNLO, and CGL 2001, respectively.
}
\label{fig:cgl_NLO_mr}
\end{figure}

To verify the valid of Eq.~(\ref{eq:mpr_NNLO}) and compare with relevant NLO formulas,
we apply Eq.~(\ref{eq:m_pi_r}) to measure the corresponding results as a reference (CGL 2001)~\footnote{
Note that the input parameters of the coupling constants: $\bar{\ell}_1$, $\bar{\ell}_2$, $\bar{\ell}_3$, $\bar{\ell}_4$,
$\tilde{r}_1$, $\tilde{r}_2$, $\tilde{r}_3$, $\tilde{r}_4$,
and $\tilde{r}_6$ are mainly quoted from Ref.~\cite{Colangelo:2001df} (CGL 2001).
},
where Eqs.~(\ref{eq:Thresholds_a}) and (\ref{eq:Thresholds_b}) are utilized to calculate
the numerical values of the $m_\pi a$ and $m_\pi^2 b$, respectively,
with the input values of $C_i's$ listed in Eq.~(\ref{eq:C123_CGL}).
The relevant results are displayed in Fig.~\ref{fig:cgl_NLO_mr}.
We also show the corresponding results directly using Eq.~(\ref{eq:mpr_NNLO}),
which are shown in Fig.~\ref{fig:cgl_NLO_mr} using cyan band as NNLO.
The corresponding NLO results are actually obtained from  the first three terms
in Eqs.~(\ref{eq:mpr_NNLO}) and (\ref{eq:mP_NNLO}), respectively~\cite{NPLQCD:2011htk}.

It is worth mentioning that three methods give similar results for pretty small $z$-values, as expected.
Moreover, it is obvious that the NNLO results are more reasonable,  and its asymptotic behaviors
are consistent with the references up to $z=7$.
Furthermore, the corresponding values at the physical point for NNLO and CGL 2001 are almost concerted,
 and difference between CGL 2001 and NLO is only about $2\%$.
As noticed in Ref.~\cite{Fu:2017apw}, it is reasonable outcome
since last three terms in Eq.~(\ref{eq:mpr_NNLO}) naturally come from
third part of $m_\pi r$ denoted in Eq.~(\ref{eq:m_pi_r}), which
absolutely dominates its behavior for large z-values~\cite{Fu:2017apw},
and is omitted in NLO.

Analogously, we discuss the validity and comparison of the Eq.~(\ref{eq:mP_NNLO}),
which are also manifested in bottom  panel of Fig.~\ref{fig:cgl_NLO_mr}.
From Eq.~(\ref{eq:m_pi_P}), it is obvious that it contains the lowest order term like $1/z$,
and highest order term like $z^6$, so it is impossible to use just a few terms
to correctly represent its $\chi$PT expressions.
However, it is very inspiring from Fig.~(\ref{fig:cgl_NLO_mr})
that the relevant asymptotic behaviors are compatible with each other.
It is pretty astonishing that the central value of NLO's results~\cite{NPLQCD:2011htk}
is more closer to that of CGL 2001.
Thus, its quantitative dependence is not clear to us and needs further investigation.

Admittedly, it is somewhat surprising that all methods
result in the pretty compatible results within statistical uncertainties.
Actually, it partially reflects the fact that LO $\chi$PT
can nicely reproduce $I=2$ $s$-wave $\pi\pi$ scattering length
with just a $0.5\%$ deviation as compared with the relevant experimental
and theoretical results~\cite{Weinberg:1966kf, Batley:2010zza, Pislak:2003sv}.
It is similar case for the slope parameters $b$ or $c$~\cite{Bijnens:1997vq,Colangelo:2001df,Bijnens:2014lea}.

We should remark that Eqs.~(\ref{eq:mpr_NNLO})
and (\ref{eq:mP_NNLO}) are valid for present lattice study ({\it i.e.}, $z<7$).
Consequently,  the extensions of NPLQCD's work~\cite{NPLQCD:2011htk} to NNLO case
is not a trivial development since it proposes a meaningful approaches
towards more sophisticated comprehension of $\pi\pi$ scattering with lattice QCD.
Meanwhile,  according to above arguments, it partially confirmed
from the perspective of theoretical analysis that the relevant NLO expressions in Ref.~\cite{NPLQCD:2011htk}
is absolutely  valid for the range of interest in this work.
Moreover, we have a limited lattice data.
In view of this,  NLO expressions~\cite{NPLQCD:2011htk} are mainly used to fit our data,
and their relevant NNLO expressions are also used to fit lattice data,
and the difference between them are regarded as systematical errors~\cite{NPLQCD:2011htk}.

To make this paper self-contained,  the essential NLO expressions in Ref.~\cite{NPLQCD:2011htk} will
be reiterated subsequently.
\begin{eqnarray}
m_\pi a &=&-\frac{z}{8\pi} + z^2 C_1 - \frac{3}{128\pi^3} z^2 \ln z, \label{eq:ma_NLO} \\
m_\pi r &=& \frac{24\pi}{z} + C_{10} + \frac{17}{6\pi}\ln z, \label{eq:mr_NLO} \\
P &=& -\frac{23\pi}{z}+ C_{13} + \frac{53}{144\pi}\ln z. \label{eq:mP_NLO}
\end{eqnarray}

\subsection{Chiral extrapolation of threshold parameters}

In the present study, lattice calculations are carried out at pion masses:
$247\,\rm MeV$, $249\,\rm MeV$,  $275\,\rm MeV$, and $384\, \rm MeV$, respectively.
In principle, we can exploit all of our data to carry out
the relevant chiral extrapolation, since Eqs.~(\ref{eq:ma_NLO}),
(\ref{eq:mr_NLO}), and (\ref{eq:mP_NLO}) just have one unknown parameters, 
and Eqs.~(\ref{eq:Thresholds_a}), (\ref{eq:mpr_NNLO}), and (\ref{eq:mP_NNLO})
contain three undetermined parameters.

\begin{figure}[!t]
\includegraphics[width=8.0cm,clip]{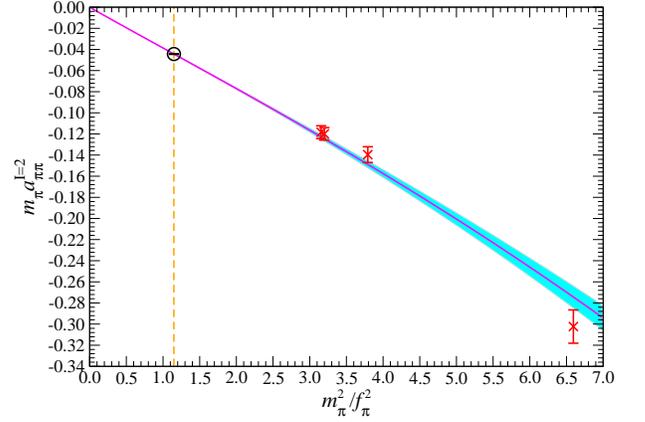}
\caption{
Chiral extrapolation of $m_\pi a $ using the lattice data
with four pion masses.
The dashed orange line indicates the physical line.
The lattice QCD + $\chi$PT prediction at physical pion mass is
the black circle on the physical line.
The shaded cyan band corresponds to statistical error, and
solid magenta curve is the central values.
}
\label{fig:maPHYS}
\end{figure}

Using  NLO $\chi$PT expressions for the scattering length $m_\pi a$ in Eq.~(\ref{eq:ma_NLO}),
the value of the constant $C_1$ can be obtained by fitting with lattice-measured $m_\pi a$
from Fit provided in Table~\ref{tab:fitstoERT}, and relevant result is also
listed in Table~\ref{tab:myFit_arp}.
Moreover, the scattering length $m_\pi a$ at physical pion mass can be predicted
using NLO $\chi$PT expressions in Eq.~(\ref{eq:mr_NLO}).
The chiral extrapolation of the scattering length $m_\pi a $ is shown in Fig.~\ref{fig:maPHYS},
and the extrapolated value at physical point is indicated by black circle
on the physical line.
The relevant results are given in Table~\ref{tab:myFit_arp} as Fit-A.

On the same time, the corresponding fitting results with NNLO $\chi$PT expression in Eq.~(\ref{eq:Thresholds_a})
are provided in Table~\ref{tab:myFit_arp}. 
In practice, as pointed out in Ref.~\cite{Yagi:2011jn}, $C_7$ in Eq.~(\ref{eq:Thresholds_a})
is hard to be determined, therefore, in this preliminary study,
it is fixed to its phenomenological value listed in Eq.~(\ref{eq:C123_CGL}).
The differences of two fits are referred to as the estimated  systematic uncertainties,
which are regarded as an important systematic error by NPLQCD
due to the effects from higher orders (NNLO) in the chiral expansion~\cite{NPLQCD:2011htk}.

\begin{table}[b!]
\caption{Results of the chiral fits for $m_\pi a$, $m_\pi r$, and $P$.
Fit-A uses NLO $\chi$PT expressions, while Fit-B adopts NNLO $\chi$PT expressions.
}
\centering
\begin{tabular*}{\linewidth}{@{\extracolsep{\fill}}lll}
\hline\hline
{\rm{Quantity}} & \rm{Fit-A} & \rm{Fit-B} \\
\hline\hline
$m_\pi a$                & $-0.04433(32)$     & $-0.04270(105)$     \\
$C_1$                    & $0.00115(24)$      & $0.00251(77)$   \\
$\chi^2/\mathrm{dof}$    & $5.8/3$            & $0.379/2$        \\
\hline
$m_\pi r$                & $57.41(1.04)$      & $52.65(3.32)$  \\
$C_{10}$                 & $-8.34(1.04)$      & $-14.27(3.28)$   \\
$\chi^2/\mathrm{dof}$    & $4.33/3$           & $2.2/2$     \\
\hline
$P$                      & $-52.79(1.09)$     & $-44.13(2.62)$  \\
$C_{13}$                 & $9.96(81))$        & $21.42(2.56)$  \\
$\chi^2/\mathrm{dof}$    & $11.4/3$           & $0.281/2$      \\
\hline\hline
\end{tabular*}
\label{tab:myFit_arp}
\end{table}

This leads to our ultimate results for scattering length $m_\pi a$ as
\begin{eqnarray}
m_\pi a &=& -0.04433(32)(163).
\label{eq:Csfit_a}
\end{eqnarray}
It is interesting to note that this systematic error can be definitely comparable with its statistical error,
which is already observed in Ref.~\cite{Yagi:2011jn}.
In Table~\ref{tab:Comp_arp}, we compare this result to these relevant results accessible in the literature.
Our lattice result of $m_\pi a$ is in good agreement with the relevant data.

%
%
\begin{table}[b!]
\centering
\caption{Summary of some lattice, experimental  and phenomenological determinations
of $I=2$ $\pi\pi$ scattering length, effective range, and shape parameter at physical point.
}
\begin{tabular*}{\linewidth}{@{\extracolsep{\fill}}lll}
\hline\hline
                                   & $m_\pi a$                & $m_\pi r$           \\
\hline\hline
Bulava2016~\cite{Bulava:2016mks}   & $-0.064(12)$             & $18.1(8.4)$         \\
CP-PACS2004~\cite{CP-PACS:2004dtj} & $-0.0431(29)$            &                     \\
Yagi2011~\cite{Yagi:2011jn}        & $-0.04410(69)(18)$       &                     \\
NPLQCD2006~\cite{Beane:2005rj}     & $-0.0426(6)(3)$          &                     \\
NPLQCD2008~\cite{Beane:2007xs}     & $-0.04330(42)$           &                     \\
NPLQCD2012~\cite{NPLQCD:2011htk}   & $-0.0417(7)$             & $72.0(5.3)$         \\
PACS-CS2014~\cite{Sasaki:2013vxa}  & $-0.04263(22)(41)$       &                     \\
ETM2010~\cite{Feng:2009ij}         & $-0.04385(28)(38)$       &                     \\
ETM2015~\cite{Helmes:2015gla}      & $-0.0442(2)(^{+4}_{-0})$ &                     \\
GWU2019~\cite{Culver:2019qtx}      & $-0.0433(2)$             &                     \\
Fischer2021~\cite{Fischer:2020jzp} & $-0.0481(86)$            & $3.9(1.1)$          \\
RBC2021~\cite{RBC:2021acc}         & $-0.055(15)(68)$         &                     \\
This work                          & $-0.04433(32)(163)$      & $57.41(1.04)(4.76)$   \\
\hline
Weinberg~\cite{Weinberg:1966kf}    & $-0.06$                  &                     \\
CGL1997~\cite{Bijnens:1997vq}      & $-0.0443$                & $59.04$             \\
CGL2001~\cite{Colangelo:2001df}    & $-0.0444(10)$            & $59.03(3.90)$       \\
CGL2012~\cite{Caprini:2011ky}      & $-0.0445(14)$            &                     \\
\hline
NA48/2~\cite{Batley:2010zza}       & $-0.0444(7)(5)(8)$       &                     \\
E865~\cite{Pislak:2003sv}          & $-0.0454(31)(10)(8) $    &                     \\
\hline\hline
\end{tabular*}
\label{tab:Comp_arp}
\end{table}

With NLO $\chi$PT expressions for the effective range $m_\pi r$ in Eq.~(\ref{eq:mr_NLO})
and the shape parameter $P$ in Eq.~(\ref{eq:mP_NLO}),
the $C_{10}$ and $C_{13}$ values can be obtained by fitting with lattice-determined $m_\pi r$ and $P$
from Fit listed in Table~\ref{tab:fitstoERT}.
Moreover, the effective range $m_\pi r$ and $P$  at physical pion mass
can be predicted by NLO $\chi$PT.
The relevant results are provided in Table~\ref{tab:myFit_arp}.
%
\begin{figure}[t!]
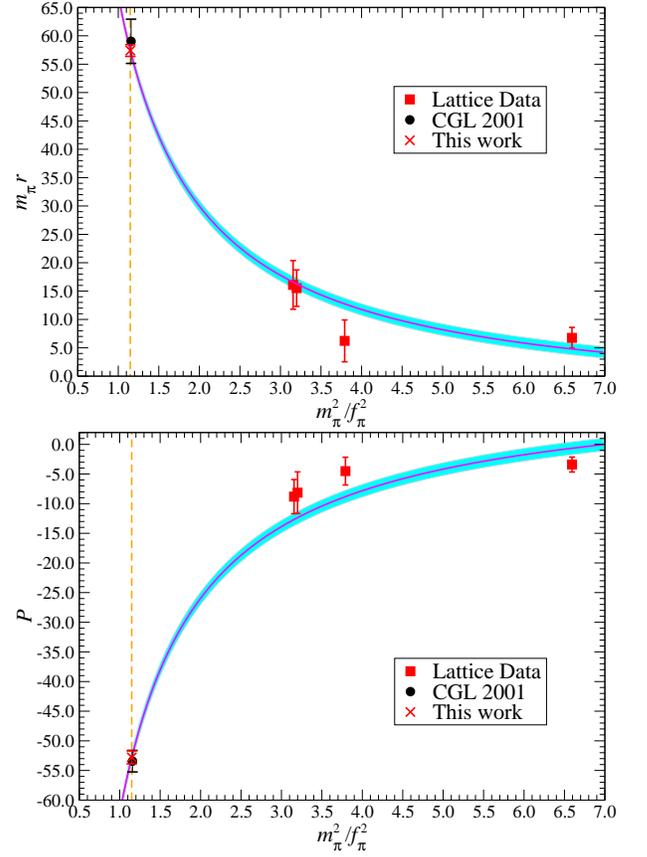

\includegraphics[width=8.0cm,clip]{mpi_mR.eps}
\includegraphics[width=8.0cm,clip]{mpi_mP.eps}
\caption{
Chiral extrapolations of $m_\pi r $ and $P$ using NLO $\chi$PT expressions.
The dashed orange line indicates the physical line.
The lattice QCD + $\chi$PT prediction at physical pion mass
is red cross on physical line,
where the statistical error and systematic  error are added in quadrature.
The Roy equation predictions~\protect\cite{Colangelo:2001df}
are the black circles on the physical line.
}
\label{fig:mrPHYS}
\end{figure}

Besides, the relevant fitting results with NNLO $\chi$PT expressions
are also given in Table~\ref{tab:myFit_arp}.
In practice, as pointed out above, $C_{12}$ in Eq.~(\ref{eq:mpr_NNLO}) and $C_{15}$ in Eq.~(\ref{eq:mP_NNLO})
are fixed to their phenomenological values listed in Eq.~(\ref{eq:C123_CGL}).
The differences of two fits are regarded as the estimated systematic errors~\cite{NPLQCD:2011htk}.
This leads to our ultimate results of $m_\pi r$ and $P$ as
\begin{eqnarray}
m_\pi r &=&  57.41(1.04)(4.76), \cr
P       &=& -52.79(1.09)(8.66).
\label{eq:Csfit_mr}
\end{eqnarray}
In Fig.~\ref{fig:mrPHYS}, the chiral extrapolations of  $m_\pi r $ and $P$
are shown in the top panel, and bottom panel, respectively.
Fig.~\ref{fig:mrPHYS} provides a comparison
of our lattice calculations and the Roy equation
values of $m_\pi r$ and $P$, which are indicated by black circles on the physical line.

It is quite inspiring that our lattice-obtained values of the scattering length $m_\pi a$ and
the effective range $m_\pi r$ are turned out
to be in reasonable accordance with the Roy equation determinations~\cite{Colangelo:2001df},
which are listed in Table~\ref{tab:Comp_arp}.
Moreover,  $P$ is also consistent with NPLQCD's lattice determination, $-75.5(16.8)$~\cite{NPLQCD:2011htk}

According to Eqs.~(\ref{eq:mr_NLO}) and (\ref{eq:mP_NLO}),
the effective range $m_\pi r$ and $P$ are divergent in the chiral limit,
which can partially interpret the extrapolated values of $m_\pi r$ and $P$
to physical point usually have relatively large statistical errors,
as compared with that of $m_\pi a$.
In practice, the statistical errors of $m_\pi r$ and $P$  are
roughly estimated by the errors of $C_{10}$ and $C_{13}$, respectively.
Moreover, they are not dependent on the pion mass (or $z$-value).
To be sure, the corresponding NNLO fits are definitely dependent on pion mass.

\subsection{Chiral extrapolation of the phase shift}
Lattice irreducible representations are carefully opted in this work to yield small scattering momenta $k$~\cite{Bulava:2016mks}.
And, $\pi\pi$ scattering phase shifts are lattice-calculated within the $t$-channel cut $k^2=m_\pi^2$.
Their lattice data of $k\cot\delta/m_\pi$ can be fit with the full formula  Eq.~(\ref{eq:pipi_nnlo}).
Due to the limited resources of computer, only $7-11$ lattice data are at disposal for each lattice ensemble.
The NNLO scattering amplitude $t(k)$ in Eq.~(\ref{eq:pipi_nnlo}) have nine independent constants,
only the relevant results of the $(0.0054, 0.018)$ ensemble can handle relatively this kind of fitting.

\begin{figure}[t!]
\includegraphics[width=8.5cm,clip]{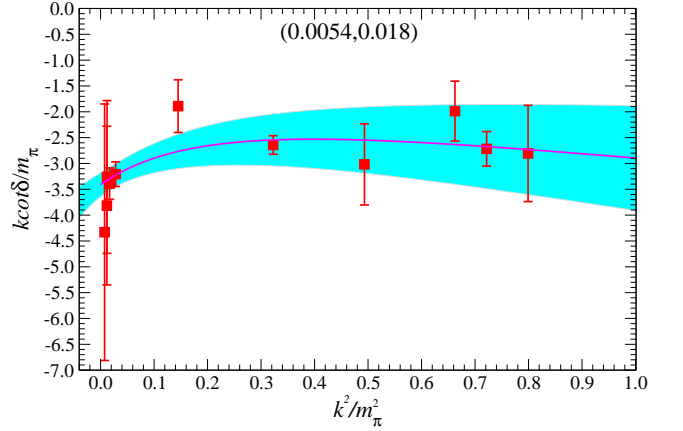}
\caption{
Three parameter fit ($C_1$, $C_{10}$, and $C_{13}$) of NLO $\chi$PT expression for $k\cot\delta/m_\pi$
to the results of the $(0.0054,0.018)$  ensemble.
The shaded cyan bands correspond to statistical error, and solid magenta curve is the central values.
}
\label{fig:pipiphasefit}
\end{figure}

As it is done in Eq.~(20) of Ref.~\cite{NPLQCD:2011htk}, the NLO $\chi$PT amplitude (one-loop level) is fitted to the
lattice results at all of the calculated energies.
And these values are exhibited in Fig.~\ref{fig:pipiphasefit} for the $(0.0054, 0.018)$ ensemble.
The shaded cyan bands correspond to statistical error, and solid magenta curve is the central values.
The fitting values for $C_1$, $C_{10}$, and $C_{13}$ are
\begin{eqnarray}
C_1^{NLO}    &=& 0.00380(30), \cr
C_{10}^{NLO} &=& -3.36(4.04), \cr
C_{13}^{NLO} &=& -14.77(9.75),
\end{eqnarray}
with a $\chi^2/{\rm dof} = 5.25/8$.

In principle, the lattice results for the $(0.0054, 0.018)$ ensemble given in Table~\ref{tab:pp_Io_kcotk}
can be fit to the NNLO $\chi$PT amplitude (two-loop level) formula
\begin{eqnarray}
\frac{k\cot{\delta}}{m_\pi} &=& \sqrt{1+\frac{k^2}{m_\pi^2}}\frac{1}{t_{\rm LO}(k)}
\left[1 - \frac{t_{\rm NLO}(k)}{t_{\rm LO}(k)}
-\frac{t_{\rm NNLO}(k)}{t_{\rm LO}(k)}\right] \cr
&&+ i\frac{k}{m_\pi},
\label{eq:chiptfitform}
\end{eqnarray}
where $t_{\rm LO}$, $t_{\rm NLO}$ and $t_{\rm NNLO}$ are the LO, NLO and NNLO contributions
to the NNLO scattering amplitude $t(k)$ in Eq.~(\ref{eq:pipi_nnlo}), and
\begin{eqnarray}
\hspace{-0.5cm}t_{\rm LO} &=&  -\frac{1}{8\pi}\hspace{-0.05cm}\left[1 + \frac{2k^2}{m_\pi^2}\right]\hspace{-0.05cm}z \cr
\hspace{-0.5cm}t_{\rm NLO} &=&  \left[C_1 + \frac{k^2}{m_\pi^2}C_2 + \frac{k^4}{m_\pi^4}C_3\right]\hspace{-0.05cm}z^2 \cr
\hspace{-0.5cm}&& -\frac{1}{4\pi^3}\left[\frac{3}{32} + \frac{5}{12}\frac{k^2}{m_\pi^2}
                  + \frac{5}{9}\frac{k^4}{m_\pi^4}\right]\hspace{-0.05cm}z^2\ln z  \cr
\hspace{-0.5cm}t_{\rm NNLO} &=& \left[C_4 + \frac{k^2}{m_\pi^2} C_5 + \frac{k^4}{m_\pi^4}C_6\right]z^3 \cr
\hspace{-0.5cm}&&+\left[C_7 + \frac{k^2}{m_\pi^2} C_8 + \frac{k^4}{m_\pi^4} C_9\right]\hspace{-0.05cm}z^3\ln z \cr
\hspace{-0.5cm}&&+\frac{1}{64\pi^5}\hspace{-0.1cm}\left[\frac{31}{24} + \frac{169}{144} \frac{k^2}{m_\pi^2}
-\frac{35}{27}\frac{k^4}{m_\pi^4}\right]\hspace{-0.05cm}z^3\hspace{-0.05cm}\left(\ln z\right)^2.
\end{eqnarray}
Note that the constants $C_2$, $C_5$, $C_8$, $C_{3}$, $C_{6}$, and $C_{9}$ can be  expressed
in Eq.~(\ref{eq:newConst258369})
by nine independent low-energy constants: $C_1$, $C_4$, $C_7$, $C_{10}$,
$C_{11}$, $C_{12}$, $C_{13}$, $C_{14}$ and $C_{15}$.

In practice, as partially pointed out in Ref.~\cite{Yagi:2011jn}, $C_7$, $C_8$ and $C_9$ in Eq.~(\ref{eq:chiptfitform})
are hard to be determined, and only eleven data are at disposal, therefore, in this preliminary study,
they are fixed to its phenomenological value listed in Eq.~(\ref{eq:C123_CGL}).
The differences of two fits are regarded as the estimated systematic errors~\cite{NPLQCD:2011htk}.
This leads to our ultimate interpolated ERE parameters:
\begin{eqnarray}
m_\pi a &=& 0.04084(39)(18), \cr
m_\pi r &=& 62.43(4.04)(1.76), \cr
P  &=&  -77.74(9.75)(13.73),
\end{eqnarray}
which are reasonably consistent within uncertanties, but less precise, than the
threshold determinations of Eq.~(\ref{eq:Csfit_a}) and Eq.~(\ref{eq:Csfit_mr}).
For a better determination of the threshold parameters from global fit,
one definitely requires more accurate lattice QCD calculations.

\section{Summary and Conclusion}
\label{sec:discussion}

In this work, by using the gauge configurations
with $N_f=3$ flavors of Asqtad-improved staggered dynamical
quarks~\cite{Aubin:2004wf,Bernard:2001av,stag_fermion},
a lattice study of $I=2$ $s$-wave $\pi\pi$ scattering
is carried out over a range of momenta below the inelastic threshold,
and total momenta $\mathbf{P}=[0,0,0]$, $[0,0,1]$, $[0,1,1]$, $[1,1,1]$,
$[0,0,2]$, $[0,2,2]$, $[0,0,3]$, $[2,2,2]$, and $[0,0,4]$,
where L\"uscher's technique~\cite{Luscher:1986pf,Luscher:1990ux,Luscher:1990ck}
is used to extract phase shifts with lattice-calculated energy-eigenstates.

The technique of the ``moving'' wall source introduced in Refs.~\cite{Kuramashi:1993ka,Fukugita:1994ve}
is exploited to calculate two quark-line diagrams for $I=2$ $\pi\pi$ scattering~\cite{Kuramashi:1993ka,Fukugita:1994ve}.
Consequently, the signals of relevant correlators are substantially improved
by comparison with our previous works~\cite{Fu:2012gf,Fu:2013ffa},
since we make use of lattice ensembles with relatively large $L$~\cite{Lepage:1989hd,Fukugita:1994ve,Fu:2016itp}.

In this work, we expand the NPLQCD's technique~\cite{NPLQCD:2011htk}
to analyze $I=2$ $\pi\pi$ scattering amplitude at NNLO in $\chi$PT,
which just adds six  more low-energy constants.
As it is shown above, it is definitely not a trivial extension
since it demonstrates some fresh characteristics, and can predict more reasonable outcomes
at least from a theoretical perspective.
It provides some analytical expressions towards the successful and sophisticate fit or global fit,
which is high desirous to adopt the corresponding NNLO $\chi$PT expression
for more sophisticated lattice study in the future.

The lattice-measured phase shifts allow us to
measure the scattering length  $m_\pi a$, effective range  $m_\pi r$, and shape parameter $P$.
The chiral extrapolations of $m_\pi a$, $m_\pi r$, and $P$  are carried out at NLO,
and its NNLO fits are regarded as the important systematic uncertainty in $\chi$PT~\cite{NPLQCD:2011htk}.
Extrapolated to the physical point, we give rise to the final outcomes as
\begin{eqnarray}
m_\pi a  &=& -0.04433(32)(163),  \cr
m_\pi r  &=&  57.41(1.04)(4.76) , \cr
P        &=& -52.79(1.09)(8.66) , \nonumber
\nonumber
\end{eqnarray}
which are all in reasonable agreement with the relevant determinations.
Note that the systematic errors are only estimated from the corresponding NNLO expressions.
We view it as one of the important results for the current investigation.

Since the systematic errors  which are estimated from the corresponding NNLO expressions
are now well under control, the lattice spacing error could be one of an important source
of the uncertainty of the relevant physical quantities~\cite{Buchoff:2008ve,Buchoff:2008hh}.
According to the discussions in Ref.~\cite{Feng:2009ij}, the lattice
spacing dependence of them is mild for the lattice calculations,
and can be hence ignored in this preliminary study.
However, one should bear in mind that this issue
should be settled in the more sophisticated lattice examination~\cite{NPLQCD:2011htk}.

For each lattice ensemble, we just calculate $7-11$  points,
simply due to the lack of computational resources,
and the robust extraction of the shape parameter $P$
definitely needs more lattice data within $k^2/m_\pi^2<0.5$ for each lattice ensemble.
Admittedly, the most efficient way to improve the statistical errors of $P$ is
working on lattice ensembles with different size $L$ for a given pion mass,
as is done for the isospin-$2$ $\pi\pi$ scattering in Ref.~\cite{NPLQCD:2011htk}.
It is  high desirable to investigate the isospin-$2$ $\pi\pi$ scattering  with large $L$(i.e., 64, etc),
so more lattice points are within $k^2/m_\pi^2<1.0$, and the relevant results are definitely more robust.

\section*{Acknowledgments}
This project was strongly and gratuitously supported by High Performance Computing Center of Sichuan University
and intensely aided by the advanced computation facility at College of physics, Sichuan University.
This work is partially supported by the National Magnetic Confinement Fusion Program of China (Grant
No. 2013GB109000), the Fundamental Research Funds for the Central Universities (Grant No. 2010SCU23002).
and the National Natural Science Foundation of China (Grant No. 12335002 and 11175124).
We would like to deliver our earnest gratitude to MILC for permitting us to use some initial gauge configurations
and the GPL open of its MILC codes without reservation.
We genuinely thank  Carleton DeTar for his copying initial gauge configurations
and instructing the necessary theoretical knowledge and computational skills for generating lattice.
We especially acknowledge Prof. Zheng Hanqing for his enlightening and constructive comments.
We would like to deliver our appreciation to many warm-hearted persons
for donating us enough removable hard drives to store the light quark propagators,
and contributing us their computation quota for this work.
We specially convey our  gratitude to the Institute of Nuclear Science and Technology, Sichuan
University, from which the computer resources and electricity costs were freely furnished.

%
%
\appendix
%
\section{$I=2$ NNLO $\pi\pi$ scattering amplitude}
\label{app:ChPT NNLO}
In this Appendix, we derive $I=2$ $\pi\pi$ partial scattering amplitude
at NNLO in $\chi$PT~\cite{Gasser:1983yg,Bijnens:1995yn,Bijnens:1997vq,Colangelo:2001df,Bijnens:2014lea,NPLQCD:2011htk}.

For elastic $\pi\pi$ scattering, the Mandelstam variables are written
in units of physical pion mass squared $m_\pi^2$  as
$$
s = 4+4\bar{k}^2, \,\,\,\,
u = -2\bar{k}^2(1+\cos\theta), \,\,\,\,
t = -2\bar{k}^2(1-\cos\theta),
$$
where $\theta$ is the scattering angle, $k$
is the CoM three-momentum of pions, and $\bar{k} \equiv k/m_\pi$.
In the $s$-channel, one evaluates $I=2$ $\pi\pi$ scattering amplitude
with combination of scattering amplitude $A$ as~\cite{Bijnens:1997vq}
\begin{equation}
T^2(s,t) = A(t,u,s) + A(u,s,t) .
\end{equation}
Using the Formula (4.14) in Ref.~\cite{Bijnens:1997vq}, we obtain
\begin{widetext}
\begin{eqnarray}
T^2(s,t) &=& x_2   \left[ -2-4\bar{k}^2 \right]
+x_2^2 \left[2b_1+b_2(t+u) + b_3 (t^2 + u^2) + b_4((u-s)^2+(s-t)^2)\right] \cr
&&{}\cr
&&+x_2^2 \left[F^{(1)}(t) + F^{(1)}(u) + G^{(1)}(t,u)  + G^{(1)}(t,s) + G^{(1)}(u,s) + G^{(1)}(u,t)\right] \cr
&&{}\cr
&&+x_2^3 \left[b_5(t^3+u^3) + b_6 ( t(u-s)^2 + u(s-t)^2\right] \cr
&&{}\cr
&&+x_2^3 \left[F^{(2)}(t) + F^{(2)}(u) + G^{(2)}(t,u)  + G^{(2)}(t,s) + G^{(2)}(u,s) + G^{(2)}(u,t)\right],
\label{eq:T1_st_def}
\end{eqnarray}
where loop integrals $F$ and $G$ are denoted in Ref.~\cite{Bijnens:1995yn,Bijnens:1997vq},
and $x_2\equiv 2m_\pi^2/f_\pi^2$.
It is easy to see that
\begin{eqnarray}
\hspace{-1.0cm}&&x_2^2\left[2b_1 + b_2(t+u) + b_3(t^2+u^2) + b_4\left((u-s)^2+(s-t)^2\right)\right]\cr
\hspace{-1.0cm}&&\hspace{3cm}=
x_2^2\left[2b_1+32b_4 + 4(-b_2+24 b_4)\bar{k}^2+ \frac{32}{3}(b_3+7b_4)\bar{k}^4\right]P_0(\cos\theta)
%
+\frac{16}{3}(b_3+b_4)\bar{k}^4 P_2(\cos\theta)\label{eq:A3_1}\\
%
\hspace{-1.0cm}&&x_2^3 \left\{b_5(t^3+u^3) + b_6\left[t(u-s)^2 + u(s-t)^2\right]\right\}
=-x_2^3 64 b_6 \bar{k}^2\left[1 + \frac{8}{3}\bar{k}^2\right]  P_0(\cos\theta)
+x_2^3 \frac{128}{3}b_6 \bar{k}^4 P_2(\cos\theta) + {\cal{O}}(\bar{k}^6) ,
\label{eq:A3_2}
\end{eqnarray}
where  the Legendre polynomials $P_0(\cos\theta)=1$ and
$P_2(\cos\theta) = 3/2\cos^2\theta-1$.
It is also easy to show that
\begin{eqnarray}
G^{(1)}(t,s)+G^{(1)}(u,s) = 2\bar{J}(s)\left[1+4\bar{k}^2 + 4\bar{k}^4 \right] P_0(\cos\theta) + {\cal{O}}(\bar{k}^6),
\label{eq:x23_b5b6}
\end{eqnarray}
where loop function $\bar{J}(s)$ is denoted in Ref.~\cite{Gasser:1983yg}.
Using the power representations of the loop integrals (B.1) in Ref.~\cite{Colangelo:2001df},
and  the trick in Appendix B of the previous work~\cite{Fu:2017apw},
it is plain to show that
\begin{eqnarray}
F^{(1)}(t)+ F^{(1)}(u) + G^{(1)}(t,u) + G^{(1)}(u,t)&=&
\left[\bar{J}(t) + \bar{J}(u)\right]
\left[\frac{11}{6} + \frac{14}{3}\bar{k}^2 + 8\bar{k}^4 + \frac{16}{3}\bar{k}^4 x^2\right] \cr
&&- \cos\theta\left[\bar{J}(t)- \bar{J}(u)\right]2\bar{k}^2\left[1+\frac{10}{3}\bar{k}^2\right]
+ {\cal{O}}(\bar{k}^6)
\label{eq:FFGG1}
\end{eqnarray}
and
\begin{eqnarray}
\bar{J}(u)+\bar{J}(t) &=&
-\frac{\bar{k}^2}{24\pi^2}\left(1-\frac{4}{15}\bar{k}^2\right)P_0(\cos\theta) +
\frac{\bar{k}^4}{180\pi^2}\left(1-\frac{6}{7}\bar{k}^2\right)P_2(\cos\theta)+ {\cal{O}}(\bar{k}^6)  \cr
%
\cos\theta\left[\bar{J}(t)-\bar{J}(u)\right] &=&
 \frac{\bar{k}^2}{72\pi^2}\left(1 - \frac{2}{5}\bar{k}^2\right)P_0(\cos\theta)+
 \frac{\bar{k}^2}{36\pi^2}\left(1 - \frac{2}{5}\bar{k}^2\right)P_2(\cos\theta)
+ {\cal{O}}(\bar{k}^6) .
\label{eq:Jutpm}
\end{eqnarray}
%
%
Plugging the Eq.~(\ref{eq:Jutpm}) into Eq.~(\ref{eq:FFGG1}),  we obtain
\begin{eqnarray}
\hspace{-0.5cm}F^{(1)}(t) + F^{(1)}(u) + G^{(1)}(t,u) + G^{(1)}(u,t) =
\left(-\frac{11}{144\pi^2} \bar{k}^2 - \frac{109}{540\pi^2} \bar{k}^4\right)P_0(\cos\theta)
-\frac{49}{1080\pi^2} \bar{k}^4 P_2(\cos\theta)+{\cal{O}}(\bar{k}^6).\nonumber
\end{eqnarray}
%
%
And the real part of $G^{(1)}(t,s)+G^{(1)}(u,s)$ is
\begin{eqnarray}
G^{(1)}(t,s)+G^{(1)}(u,s) &=&
\frac{1}{4\pi^2}\hspace{-0.1cm}\left[ 1+3\bar{k}^2 + \frac{2}{3}\bar{k}^4\right]\hspace{-0.1cm} P_0(\cos\theta)
+ {\cal{O}}(\bar{k}^6).
\label{eq:Guts1}
\end{eqnarray}
%
Consequently, we get
\begin{eqnarray}
F^{(1)}(t)+F^{(1)}(u)+G^{(1)}(t,u)+G^{(1)}(t,s)+G^{(1)}(u,s) +G^{(1)}(u,t)
&=&\left[\frac{1}{4\pi^2}+\frac{97}{144\pi^2}\bar{k}^2 - \frac{19}{540\pi^2}\bar{k}^4\right]P_0(\cos\theta)\cr
&&-\frac{49}{1080\pi^2}\bar{k}^4 P_2(\cos\theta)+ {\cal{O}}(\bar{k}^6).
\end{eqnarray}
%
The scattering amplitude at NLO is gotten as
\begin{eqnarray}
T^2_{NLO}(s,t) &=& x_2 \left\{-2-\bar{k}^2 + x_2 \left[\frac{1}{4\pi^2} + 2b_1 + 32 b_4 \right] +
x_2\left[\frac{97}{144\pi^2}-4b_2+ 96b_4 \right] \bar{k}^2 +
x_2  \left[ -\frac{19}{540\pi^2} + \frac{32}{3}\left(b_3 + 7b_4\right)\right]\bar{k}^4 \right\} \cr 
&&+
x_2^2 \left[ -\frac{49}{1080\pi^2} +\frac{16}{3}\left(b_3+b_4\right)\right]  \bar{k}^4 P_2(\cos\theta)
+ {\cal{O}}(\bar{k}^6) ,
\label{eq:NLO_I1}
\end{eqnarray}
which is consistent with relevant results in Ref.~\cite{NPLQCD:2011htk}.
Now we begin to calculate NNLO term.
It is easy to verify that
\begin{eqnarray}
&&G^{(2)}(t,s) + G^{(2)}(u,s) = \cr
&&\hspace{1.24cm}\bar{J}(s)\left[
-\frac{387}{216\pi^2}-4b_1-64b_4 +
\left(-\frac{82}{9\pi^2}-8b_1 + 8b_2 -320 b_4 \right)\bar{k}^2+
\left(-\frac{431}{27\pi^2} + 16b_2 - \frac{64}{3}b_3 - \frac{1600}{3} b_4 \right)\bar{k}^4
\right]+\cr
&&\hspace{1.0cm}K_1(s)\left[
-3-\frac{62}{3}\bar{k}^2 -\frac{128}{3}\bar{k}^4
  +\frac{1}{12}\pi^2\bar{k}^4 + \frac{1}{9}\pi^2\bar{k}^2\right]
+K_2(s)\left[\frac{2}{9} - \frac{16}{9}\bar{k}^2
              -\frac{1}{9}\pi^2-\frac{7}{36}\pi^2\bar{k}^2-\frac{1}{12}\pi^2\bar{k}^4\right] + \cr
&&
\hspace{1.0cm}K_3(s)\left[\frac{26}{9} + \frac{76}{9}\bar{k}^2 + \frac{16}{3}\bar{k}^4 \right]
 + {\cal{O}}(\bar{k}^6).
\label{eq:G_tu_s}
\end{eqnarray}
Using the Taylor series expansions~(B.1) in Ref~\cite{Colangelo:2001df},
we can straightforwardly calculate each term of Eq.~(\ref{eq:G_tu_s}),
and it is easy to verify that
\begin{eqnarray}
G^{(2)}(t,s) + G^{(2)}(u,s) &=& \frac{1}{8\pi^2}
\bigg\{ -\frac{131}{72\pi^2} + \frac{11}{72}  - 4b_1 - 64b_4 +
\left(-\frac{551}{72\pi^2} + \frac{635}{864} -4b_1 + 8b_2 -256b_4 \right)\bar{k}^2 \cr
&&{ }\cr
&&+\left(-\frac{1169}{108\pi^2} + \frac{1946}{1440} +  \frac{16}{3}b_1 + 8b_2
                        - \frac{64}{3} b_3 - 256 b_4 \right)\bar{k}^4
\bigg\} P_0(\cos\theta)  + {\cal{O}}(\bar{k}^6).
\label{eq:Gtus_2}
\end{eqnarray}
%
%
Following the notations in Refs.~\cite{Bijnens:1995yn,Bijnens:1997vq},
and using the trick in our previous work in Ref.~\cite{Fu:2017apw}, we can write
$F^{(2)}(t) + G^{(2)}(u, t)  + F^{(2)}(u) + G^{(2)}(t, u)$ as an elegant form
\begin{eqnarray}
&& F^{(2)}(t) + G^{(2)}(u, t)  + F^{(2)}(u) + G^{(2)}(t, u) \cr
&&=\hspace{-0.05cm}P_0(x) \left[\bar{J}(u) + \bar{J}(t)\right]
\left[-\frac{1}{16\pi^2} \left(\frac{8}{27} + \frac{553}{18} \bar{k}^2 \right)\right] +
x\left[\bar{J}(t) - \bar{J}(u)\right]
\frac{\bar{k}^2}{16\pi^2}\left(\frac{85}{2}+\frac{754}{9}\bar{k}^2\right)\cr
&&{ }\cr
%
&&+P_0(x) \left[\bar{J}(u)+\bar{J}(t)\right] \left(-1-6\bar{k}^2\right)b_1 +
x\left[\bar{J}(t)-\bar{J}(u)\right] 6\bar{k}^2 b_1 \cr
&&{ }\cr
&&+P_0(x) \left[\bar{J}(u)+\bar{J}(t)\right] \frac{b_2}{3}\left(10\bar{k}^2 + 8\right)+
x\left[\bar{J}(t)-\bar{J}(u)\right]\frac{b_2}{3}\left(6-40\bar{k}^2\right)\bar{k}^2  \cr
&&{ }\cr
&&+P_0(x)\left[\bar{J}(u)+\bar{J}(t)\right]\frac{b_3}{6}\left(96\bar{k}^2 +64\right) +
x\left[\bar{J}(t)-\bar{J}(u)\right]\frac{b_3}{6}\left( 32 + 80 \bar{k}^2\right)\bar{k}^2 \cr
&&{ }\cr
&&+P_0(x)\left[\bar{J}(u)+\bar{J}(t)\right] \frac{b_4}{6}\left(- 448\bar{k}^2 -96\right)+
x\left[\bar{J}(t)-\bar{J}(u)\right]\frac{b_4}{6}\left( 320 + 816 \bar{k}^2\right)\bar{k}^2 \cr
&&{ }\cr
&&+P_0(x)\left[K_1(u)+ K_1(t)\right]
\frac{1}{36}\left(-96 + \frac{5\pi^2}{2} - 486\bar{k}^2 - \frac{\pi^2}{4}\bar{k}^2 \right)  +
x\left[K_1(t)- K_1(u)\right]\frac{1}{36}\left(454 - \frac{11\pi^2}{4} + 1328 \bar{k}^2\right)\bar{k}^2 \cr
&&{ }\cr
&&+P_0(x)\left[K_2(u) + K_2(t)\right]
\frac{1}{288}\left(- 10\pi^2 \bar{k}^2 -256 \bar{k}^2 - 184\right) +
x\left[K_2(t) - K_2(u)\right]\frac{1}{288}\left(10\pi^2  - 256\right)\bar{k}^2 \cr
&&{ }\cr
&&+P_0(x)\left[K_3(u) + K_3(t)\right]\frac{1}{9}\left(54 \bar{k}^2 + 10\right) +
x\left[K_3(t) - K_3(u)\right]\frac{1}{9}\left( -50 -48\bar{k}^2 \right)\bar{k}^2 \cr
&&{ }\cr
&&+P_0(x)\left[K_4(u) + K_4(t)\right]\left(\frac{20}{3} + 10\bar{k}^2 \right) +
x\left[K_4(t) - K_4(u)\right]\frac{10}{3} \bar{k}^2.
\label{EQ:FGFG2}
\end{eqnarray}
It is ready to show that
\begin{eqnarray}
K_1(u) + K_1(t) &=&
\frac{\bar{k}^2}{64\pi^4} \left(1-\frac{2}{9}\bar{k}^2+\frac{4}{45}\bar{k}^4\right) P_0(\cos\theta)
-\frac{\bar{k}^4}{576\pi^4}\left(1-\frac{4}{5} \bar{k}^2 \right)P_2(\cos\theta) + {\cal{O}}(\bar{k}^6) \cr
%
\cos\theta\left[K_1(t) - K_1(u)\right] &=&
- \frac{\bar{k}^2}{192\pi^4}\left[1 - \frac{1}{3}\bar{k}^2\right]P_0(\cos\theta)
- \frac{\bar{k}^2}{96 \pi^4}\left[1 - \frac{1}{3}\bar{k}^2\right]P_2(\cos\theta)
+ {\cal{O}}(\bar{k}^6),
\end{eqnarray}
and similar for $K_2$, $K_3$, and $K_4$.
After some algebraic manipulations, and considering Eq.~(\ref{eq:Gtus_2}), we finally get
\begin{eqnarray}
&&F^{(2)}(t) + F^{(2)}(u) + G^{(2)}(t,u)  + G^{(2)}(t,s) + G^{(2)}(u,s) + G^{(2)}(u,t)= \cr
&&P_0(\cos\theta) \Bigg\{ \frac{1}{8\pi^2}\left[-\frac{1}{16\pi^2}\frac{262}{9}
\hspace{-0.05cm}+\hspace{-0.05cm}\frac{1}{16\pi^2}\frac{22\pi^2}{9}\hspace{-0.05cm}-\hspace{-0.05cm}4b_1
\hspace{-0.05cm}-\hspace{-0.05cm}64 b_4\right]
- \frac{\bar{k}^2}{8\pi^2}\left[\frac{1}{16\pi^2}\frac{10591}{81}
\hspace{-0.05cm}-\hspace{-0.05cm} \frac{1}{16\pi^2}\frac{145\pi^2}{12}
\hspace{-0.05cm}+\hspace{-0.05cm} \frac{11}{3}b_1\hspace{-0.05cm} -\hspace{-0.05cm} \frac{64}{9}b_2
\hspace{-0.05cm}+\hspace{-0.05cm} \frac{32}{9}b_3\hspace{-0.05cm} +\hspace{-0.05cm} \frac{752}{3}b_4 \right] \cr
&&\hspace{1.2cm} + \frac{\bar{k}^4}{3\pi^2}
                   \left[ -\frac{75997}{17280\pi^2} + \frac{16649}{34560}
                     + \frac{89}{30}b_1   + \frac{124}{45}b_2 - \frac{424}{45}b_3 - \frac{3824}{45}b_4
                     \right] + {\cal{O}}(\bar{k}^6)
\Bigg\} + \cr
&& P_2(\cos\theta) \frac{ \bar{k}^4}{3\pi^2}
\left[ - \frac{1}{16\pi^2}\frac{67}{2160} -\frac{1}{16\pi^2}\frac{127\pi^2}{432}
+ \frac{29}{60} b_1 + \frac{19}{90}b_2 + \frac{28}{45} b_3 + \frac{188}{45} b_4  \right] + {\cal{O}}(\bar{k}^8) .
\label{eq:FFGGGG2}
\end{eqnarray}
Considering the NLO $\pi\pi$ amplitude in Eq.~(\ref{eq:NLO_I1}),
and plugging Eq.~(\ref{eq:FFGGGG2}) and Eq.~(\ref{eq:A3_2}) into Eq.~(\ref{eq:T1_st_def}),
we get its NNLO expression
\begin{eqnarray}
T^2(s,t) &=&
P_0(\cos\theta) \Bigg\{ -2x_2\left[ 1 - \frac{x_2}{16\pi^2}\left(2 + \bar{b}_1 + 16\bar{b}_4 \right)
+\frac{x_2^2}{256\pi^4}\left(\frac{262}{9}
- \frac{22\pi^2}{9} +  4\bar{b}_1 + 64 \bar{b}_4 \right)\right] \cr
&-&4x_2 \bar{k}^2\left[1 - \frac{x_2}{32\pi^2}\left(\frac{97}{18}-2\bar{b}_2 + 48\bar{b}_4 \right)
+\frac{x_2^2}{512\pi^4}\left(\frac{10591}{81} - \frac{145\pi^2}{12}
+ \frac{11}{3}\bar{b}_1 - \frac{64}{9}\bar{b}_2 + \frac{32}{9}\bar{b}_3
+ \frac{752}{3}\bar{b}_4 + 32\bar{b}_6 \right)\right]\cr
&+& \frac{2x_2^2}{3\pi^2} \bar{k}^4 \left[
-\frac{19}{360} + \bar{b}_3+7\bar{b}_4
+\frac{x_2}{16\pi^2}
\left( -\frac{75997}{2160\pi^2} + \frac{16649\pi^2}{4320}
+ \frac{89}{60}\bar{b}_1   + \frac{62}{45}\bar{b}_2 - \frac{212}{45}\bar{b}_3
- \frac{1912}{45}\bar{b}_4 - 16\bar{b}_6  \right)
\right]
\Bigg\}  \cr
\hspace{-0.3cm}&+&\hspace{-0.05cm} P_2(\cos\theta)\bar{k}^4 \left[
\frac{x_2^2}{3\pi^2}\hspace{-0.1cm} \left( -\frac{49}{360}
\hspace{-0.05cm} + \hspace{-0.05cm} \bar{b}_3 \hspace{-0.05cm}+\hspace{-0.05cm} \bar{b}_4 \right)
+ \frac{x_2^3}{48\pi^4}\hspace{-0.1cm}\left( - \frac{67}{2160} \hspace{-0.05cm}-\hspace{-0.05cm}\frac{127\pi^2}{432}
\hspace{-0.05cm}+\hspace{-0.05cm} \frac{29}{60} \bar{b}_1
\hspace{-0.05cm}+\hspace{-0.05cm} \frac{19}{90}\bar{b}_2
\hspace{-0.05cm}+\hspace{-0.05cm} \frac{28}{45} \bar{b}_3
\hspace{-0.05cm}+\hspace{-0.05cm} \frac{188}{45} \bar{b}_4
\hspace{-0.05cm}+\hspace{-0.05cm} 8 \bar{b}_6 \right)  \right] .
\label{app:T_11}
\end{eqnarray}
It is traditional to expand the combination with the isospin-$2$ in the $s$-channel
$T^2(s,t)$ into the partial waves,
\begin{equation}
T^2(s,t) = 32\pi\sum_{\ell=0}^{\infty}(2\ell+1)P_\ell(\cos\theta)t_{\ell}^{2}(s) .
\end{equation}
Consequently, we get the partial wave for $s$-wave ($\ell=0$)
\begin{eqnarray}
\hspace{-1.0cm}t_0^2(k) &=&
-\frac{m_\pi^2}{8\pi f_\pi^2}\left[ 1 - x\left(2 + \bar{b}_1+ 16\bar{b}_4 \right)
+ x^2\left(\frac{262}{9}-\frac{22\pi^2}{9}+4\bar{b}_1+64\bar{b}_4 \right)\right] \cr
\hspace{-1.0cm}&&-\frac{1}{4\pi f_\pi^2}\left[1 - \frac{x}{2}\left(\frac{97}{18}-2\bar{b}_2+48\bar{b}_4\right)
         +\frac{x^2}{2}\left(\frac{10591}{81} - \frac{145\pi^2}{12}
+ \frac{11}{3}\bar{b}_1 - \frac{64}{9}\bar{b}_2 + \frac{32}{9}\bar{b}_3
+ \frac{752}{3}\bar{b}_4 + 32\bar{b}_6 \right)\right]k^2\cr
\hspace{-1.0cm}&&+ \frac{1}{12\pi^3 f_\pi^4} \left[
-\frac{19}{360} + \bar{b}_3+7\bar{b}_4
+x \left( -\frac{75997}{2160} + \frac{16649\pi^2}{4320}
+ \frac{89}{60}\bar{b}_1   + \frac{62}{45}\bar{b}_2 - \frac{212}{45}\bar{b}_3
- \frac{1912}{45}\bar{b}_4 - 16\bar{b}_6 \right)
\right]k^4 \cr
&& +{\cal{O}}(k^6),
\label{app:T02_amp}
\end{eqnarray}
where {$\displaystyle x=m_\pi^2/(8\pi^2 f_\pi^2)$}, and the partial wave for $D$-wave ($\ell=2$)
\begin{equation}
\frac{t_2^2(k)}{k^4} =
\frac{m_\pi^4}{120\pi^3 f_\pi^4}
\left\{-\frac{49}{360} +  \bar{b}_3 + \bar{b}_4
+ x \left[ -\frac{67}{2160} - \frac{127\pi^2}{432}
+ \frac{29}{60} \bar{b}_1 + \frac{19}{90}\bar{b}_2 + \frac{28}{45} \bar{b}_3 + \frac{188}{45}\bar{b}_4
+  8\bar{b}_6 \right]
+ {\cal{O}}(k^2)  \right\}.
\label{app:T13_amp}
\end{equation}
The near threshold behavior of partial wave amplitude $t_\ell^I(k)$
is normally expressed as a power-series expansion in $k$
\begin{equation}
\mbox{Re}\; t_\ell^I(k)\ = k^{2\ell}\left\{a_\ell^I +k^2\;b_\ell^I + k^4\;c_\ell^I + {\cal O}(k^6)\right\},
\label{app:threshexp}
\end{equation}
where the threshold parameters $a_\ell^I$, $b_\ell^I$ and $c_\ell^I$ are referred to as the scattering length,
slope parameter, and an another slope parameter, respectively.
Matching Eq.~(\ref{app:threshexp}) to  Eq.~(\ref{app:T02_amp}) and Eq.~(\ref{app:T13_amp})  yields
\begin{eqnarray}
\hspace{-1.0cm}         a_0^2 &=&
-\frac{m_\pi^2}{8\pi f_\pi^2}\left\{ 1 - x\left[2 + \bar{b}_1+ 16\bar{b}_4 \right]
+ x^2\left[\frac{262}{9}-\frac{22\pi^2}{9}+4\bar{b}_1+64 \bar{b}_4 \right]\right\}  \label{app:nnlo_a02} \\
%
\hspace{-1.0cm}m_\pi^2 b_0^2 &=&
-\frac{m_\pi^2}{4\pi f_\pi^2}
\left\{ 1 - \frac{x}{2}\left[ \frac{97}{18} - 2\bar{b}_2 + 48 \bar{b}_4 \right]
          + \frac{x^2}{2} \left[\frac{10591}{81} - \frac{145\pi^2}{12} + \frac{11}{3}\bar{b}_1
          - \frac{64}{9}\bar{b}_2 + \frac{32}{9}\bar{b}_3 + \frac{752}{3} \bar{b}_4  + 32 \bar{b}_6\right] \right\} \label{app:nnlo_b02}\\
\cr
\hspace{-1.0cm}m_\pi^4 c_0^2 &=&  \frac{m_\pi^4}{12\pi^3 f_\pi^4} \left\{
-\frac{19}{360} + \bar{b}_3+7\bar{b}_4
+x\left[ -\frac{75997}{2160} + \frac{16649\pi^2}{4320}
+ \frac{89}{60}\bar{b}_1   + \frac{62}{45}\bar{b}_2 - \frac{212}{45}\bar{b}_3
- \frac{1912}{45}\bar{b}_4 - 16\bar{b}_6  \right]
\right\}  \label{app:nnlo_c02} \\
\hspace{-1.0cm}m_\pi^4 a_2^2 &=& \frac{m_\pi^4}{120\pi^3 f_\pi^4}
\left\{-\frac{49}{360} +  \bar{b}_3 + \bar{b}_4
+ x \left[ -\frac{67}{2160} - \frac{127\pi^2}{432}
+ \frac{29}{60} \bar{b}_1 + \frac{19}{90}\bar{b}_2 + \frac{28}{45} \bar{b}_3 + \frac{188}{45}\bar{b}_4
+  8\bar{b}_6 \right]
+ {\cal{O}}(k^2)  \right\}.
\label{app:nnlo_a22}
\end{eqnarray}
\end{widetext}

It is evident that we neatly reproduce the corresponding results
of the scattering length $a_0^2$  and slope parameter $b_0^2$ at NNLO
in Ref.~\cite{Bijnens:1997vq}, as expected.
With the CGL language~\cite{Bijnens:1997vq}, the explicit formula of the slope parameter $c_0^2$
is originally rendered in Eq.~(\ref{app:nnlo_c02}).
In this Appendix, to double-check our results for the $s$-wave(especially for $k^6$ terms),
the scattering length $a_2^2$ for $D$-wave is also given in Eq.~(\ref{app:nnlo_a22}),
since the equation~(\ref{app:T_11}) contains all the information.

%
%
\section{NNLO $\chi$PT expressions}
\label{app:C NNLO}
The scattering length $a$  and slope parameter $b$ at NNLO are given in Ref.~\cite{Bijnens:1997vq},
and slope parameter $c \equiv c_0^2$ is rendered in Eq.~(\ref{app:nnlo_c02}) at NNLO,
where $\bar{b}_i$'s are quantities introduced in Refs.~\cite{Bijnens:1997vq,Colangelo:2001df}.
Plugging $\bar{b}_i$'s expressions in Eq.~(B.3) of Ref.~\cite{Colangelo:2001df}
into Eqs.~(\ref{app:nnlo_a02},\ref{app:nnlo_b02},\ref{app:nnlo_c02}),
choosing the scale $\mu$ to be the physical pion decay constant $f_{\pi,\rm phy}$,
and using the strategies in Ref.~\cite{Yagi:2011jn}, we finally obtain
\begin{widetext}
\begin{eqnarray}
m_{\pi} a &=& -\frac{z}{8\pi} + z^2 C_1 - \frac{3}{128\pi^3} z^2 \ln z
+ z^3 C_4 + z^3\ln z C_7 + \frac{31}{3072\pi^5}z^3(\ln z)^2, \label{app_C9_a} \\
m_{\pi}^2b&=&-\frac{z}{4\pi} + z^2C_2 -\frac{5}{48\pi^3}z^2\ln z
+ z^3 C_5 + C_8 z^3\ln z + \frac{169}{9216\pi^5}z^3(\ln z)^2, \label{app_C9_b} \\
m_{\pi}^4 c &=& z^2C_3 -\frac{5}{36\pi^3}z^2 \ln z + z^3 C_6 + C_9 z^3\ln z -\frac{35}{1728\pi^5}z^3(\ln z)^2,
\label{app_C9_c}
\end{eqnarray}
with
\begin{eqnarray}
\hspace{-0.5cm}C_1 &=&
\frac{1}{128\pi^3}\left[\frac{8}{3}\tilde{\ell}_{1} +\frac{16}{3}\tilde{\ell}_{2}
-\tilde{\ell}_{3} -4 \tilde{\ell}_{4} +1 \right], \label{app_ChP_C1}\cr
\hspace{-0.5cm}C_2 &=&
\frac{1}{64\pi^3}\left[\frac{8}{3}\tilde{\ell}_1 + 8\tilde{\ell}_2-4\tilde{\ell}_4
- \frac{5}{6}\right],\label{app_ChP_C2}\cr
\hspace{-0.5cm}C_3 &=&
\frac{1}{36\pi^3}\left[\tilde{\ell}_1 + 4\tilde{\ell}_2 -\frac{193}{40}\right],\label{app_ChP_C3}\cr
\hspace{-0.5cm}C_4 &=&
\frac{1}{512\pi^5}\left[\frac{16}{3}\tilde{\ell}_{1}\tilde{\ell}_{4} + \frac{32}{3}\tilde{\ell}_{2}\tilde{\ell}_{4}
-3\tilde{\ell}_{3}\tilde{\ell}_{4} -5\tilde{\ell}_{4}^2 - \frac{1}{2}\tilde{\ell}_3^{\ 2}-\frac{4}{3}\tilde{\ell}_{1} - \frac{16}{3}\tilde{\ell}_{2}-\frac{7}{4}\tilde{\ell}_{3} - \tilde{\ell}_{4} -{\frac{163}{16}}
+\frac{22}{9}\pi^2 + \tilde{r}_1 + 16 \tilde{r}_4 \right],\label{app_ChP_C4}\cr
\hspace{-0.5cm}C_5 &=&
\frac{1}{256\pi^5}\left[\frac{16}{3}\tilde{\ell}_1\tilde{\ell}_4 + 16\tilde{\ell}_2\tilde{\ell}_4
-\tilde{\ell}_3 \tilde{\ell}_4 - 5\tilde{\ell}_4^{2} + \frac{46}{27}\tilde{\ell}_1 - \frac{170}{27}\tilde{\ell}_2
-\frac{43}{12}\tilde{\ell}_3 - \frac{25}{3}\tilde{\ell}_4 + \frac{317\pi^2}{54} - \frac{13481}{648} - \tilde{r}_2
+ 24\tilde{r}_4 - 16\tilde{r}_6 \right],\label{app_ChP_C5}\cr
\hspace{-0.5cm}C_6 &=&
\frac{1}{96\pi^5}\left[\tilde{r}_3 + 7\tilde{r}_4 -16\tilde{r}_6+\frac{4}{3}\tilde{\ell}_1\tilde{\ell}_4+ \frac{16}{3}\tilde{\ell}_2\tilde{\ell}_4+\frac{21}{5}\tilde{\ell}_1 + \frac{116}{45}\tilde{\ell}_2
-\frac{89}{120}\tilde{\ell}_3-\frac{293}{30}\tilde{\ell}_4+\frac{4453\pi^2}{1440}
-\frac{5635}{1296} \right],\label{app_ChP_C6}\cr
\hspace{-0.5cm}C_7 &=&
\frac{1}{512\pi^5}\left[ -\frac{4}{3}\tilde{\ell}_{1}-8\tilde{\ell}_{2} {+} \tilde{\ell}_{3}-2\tilde{\ell}_{4}+\frac{155}{12}\right],
\label{app_ChP_C7}\cr
\hspace{-0.5cm}C_8 &=&
\frac{1}{512\pi^5} \left[ \frac{80}{9}\tilde{\ell}_1-4\tilde{\ell}_2
+5\tilde{\ell}_3 - \frac{56}{3}\tilde{\ell}_4 + \frac{3041}{54}\right],\label{app_ChP_C8}\cr
\hspace{-0.4cm}C_9 &=&
\frac{1}{96\pi^5}\left[ \frac{41}{9}\tilde{\ell}_1 + 6\tilde{\ell}_2 -\frac{20}{3}\tilde{\ell}_4+\frac{9349}{1080}\right],
\label{app_ChP_C9}
\end{eqnarray}
where  $z\equiv m_\pi^2/f_\pi^2$, and the constants $C_1$, $C_4$, and $C_7$ are actually the variants of $\ell_{\pi\pi}^{(1)}$, $\ell_{\pi\pi}^{(2)}$, and $\ell_{\pi\pi}^{(3)}$ denoted in Ref.~\cite{Yagi:2011jn}, respectively. It should be worthwhile to stress that 
the expressions are actually expanded with chiral expansion parameter,
$\xi \equiv m_\pi^2/(8\pi^2 f_\pi^2)$,
and the relevant results are rearranged in the order of $\xi$~\cite{Yagi:2011jn}.
In fact, we here just rewrite them in NPLQCD's style~\cite{NPLQCD:2011htk}.

Note that all the low-energy constants are at the scale of
$f_{\pi,\rm phy}$ in favor of the NPLQCD's notations~\cite{NPLQCD:2011htk},
and are independent of the quark mass, therefore, we can make use of them
as fitting parameters in the corresponding chiral extrapolations.
For this scale, it should be noticed that~\cite{Yagi:2011jn}
\begin{eqnarray}
\tilde{L}(\mu=f_{\pi,\rm phy}) &=&
-\ln z -\ln{\frac{f_\pi^2}{f_{\pi,\rm phy}^2}} = -\ln z - \frac{z}{4\pi^2} \bar{\ell}_4 +{\cal O}(z^2)
= -\ln z - \frac{z}{4\pi^2} \tilde{\ell}_4(\mu= f_{\pi,\rm phy}) + \frac{z}{4\pi^2} \ln z +{\cal O}(z^2). \nonumber
\end{eqnarray}

Substituting the Eqs.~(\ref{eq:Thresholds_a})-(\ref{eq:Thresholds_c})] into
the Eqs.~(\ref{eq:m_pi_r})-(\ref{eq:m_pi_P}), realigning the relevant results
in the order of $z$, after some strenuous algebraic manipulations,
it is straightforward to get NNLO $\chi$PT descriptions for effective range approximation parameters:
\begin{eqnarray}
m_\pi r &=& \frac{24\pi}{z} +64\pi^2\left(7C_1 - 2C_2\right) + \frac{17}{6\pi} \ln z
+ z\left[\frac{1}{4\pi} + 64\pi^2\left( 7C_4 -  2C_5- 32\pi C_1C_2+ 88\pi C_1^2\right)\right] \cr
&&+ z\ln z\left(448\pi^2 C_7  - \frac{152}{3} C_1 + 48 C_2 - 128\pi^2 C_8\right)
  +\frac{77}{288\pi^3} z (\ln z)^2, \label{app_ChPT_mpir}\\
P&=& -\frac{23\pi}{z}+  8\pi^2\left[28C_2-79C_1-8C_3 \right] + \frac{53}{144\pi}\ln z
\cr
&& + z\bigg[\frac{3}{16\pi}+
8\pi^2\left(-79C_4 + 28 C_5 -8 C_6 -1336\pi C_1^2
+704\pi C_1C_2 - 64\pi C_2^2 -128\pi C_1C_3\right)\bigg]\cr
&&+ z \ln z \bigg[24C_3 + 224\pi^2 C_8 - 632\pi^2 C_7 - 64\pi^2 C_9
+\frac{509}{9}C_1-\frac{76}{3}C_2\bigg]- \frac{13711}{6912\pi^3} z (\ln z)^2, \label{app_ChPT_P}
\end{eqnarray}
where just six leading terms for each equation are kept, and it is enough for the current study.
For $z-$values $<7$, using our estimated phenomenological values $C_i$'s in Eq.~(\ref{eq:C123_CGL})
via the published values of scale independent couplings $\bar{\ell}_i$ and $\tilde{r}_i$  in Refs.~\cite{Bijnens:1997vq,Colangelo:2001df,Bijnens:2014lea},
we verify that the cross terms (like $C_1^2, C_1C_2$. etc)
are with a quite limited contribution as compared with other terms, moreover, these cross terms
are usually partially cancelled each other out.

It should be worthwhile to stress that the scale dependence of $\bar{r}_i(\mu)$
is fixed with the requirement $\mu\left(d\bar{b}_i/d\mu\right)=0$,
such that $\bar{b}_i$ has no scale dependence~\cite{Bijnens:1997vq,Colangelo:2001df,Bijnens:2014lea}. Therefore,
the right hand sides of Eqs.~(\ref{eq:Thresholds_a})-(\ref{eq:Thresholds_c}) are scale independent,
though truncated at the third order of $z$~\cite{Yagi:2011jn}.
Consequently, the right hand side of Eq.~(\ref{app_C9_a}), Eq.~(\ref{app_C9_b}), and Eq.~(\ref{app_C9_c})
as a whole is likewise scale independent~\cite{Yagi:2011jn}.
Generally speaking, the right hand side of Eq.~(\ref{app_ChPT_mpir}) and Eq.~(\ref{app_ChPT_P})
are accordingly scale independent.
As a result, one can arbitrarily select the scale $\mu$.
In practice, we actually also get all the above expressions at the scale $\mu^2 = 8\pi^2 f_{\pi,\rm phy}^2$ as it done
in Ref.~\cite{Yagi:2011jn}:
\begin{eqnarray}
m_{\pi} a &=&
-\frac{z}{8\pi} + z^2 C_1 - \frac{3}{128\pi^3} z^2 \ln\left(\frac{z}{8\pi^2}\right)
+ z^3 C_4 + z^3\ln\left(\frac{z}{8\pi^2}\right)C_7
+ \frac{31}{3072\pi^5}z^3\left[\ln\left(\frac{z}{8\pi^2}\right)\right]^2, \label{app_C98_a} \\
m_{\pi}^2b &=&
-\frac{z}{4\pi} + z^2C_2 -\frac{5}{48\pi^3}z^2\ln \frac{z}{8\pi^2}
+ z^3 C_5 + C_8 z^3\ln\left(\frac{z}{8\pi^2}\right)
+ \frac{169}{9216\pi^5}z^3\left[\ln\left( \frac{z}{8\pi^2}\right)\right]^2, \label{app_C98_b} \\
m_{\pi}^4 c &=&
z^2C_3 -\frac{5}{36\pi^3}z^2\ln\left(\frac{z}{8\pi^2}\right) + z^3 C_6 + C_9 z^3\ln\left(\frac{z}{8\pi^2}\right)
 -\frac{35}{1728\pi^5}z^3\left[\ln\left(\frac{z}{8\pi^2}\right)\right]^2 \label{app_C98_c}, \\
m_\pi r &=& \frac{24\pi}{z} +64\pi^2\left(7C_1 - 2C_2\right) + \frac{17}{6\pi} \ln\left(\frac{z}{8\pi^2}\right)
+ z\left[\frac{1}{4\pi} + 64\pi^2\left( 7C_4 -  2C_5- 32\pi C_1C_2+ 88\pi C_1^2\right)\right] \cr
&&+ z\ln\left(\frac{z}{8\pi^2}\right)\left[448\pi^2 C_7  - \frac{152}{3} C_1 + 48 C_2 - 128\pi^2 C_8\right]
  +\frac{77}{288\pi^3} z\left[\ln\left(\frac{z}{8\pi^2}\right)\right]^2, \label{app_ChPT8_mpir}\\
P&=& -\frac{23\pi}{z}+  8\pi^2\left[28C_2-79C_1-8C_3 \right] + \frac{53}{144\pi}\ln\left(\frac{z}{8\pi^2}\right)\cr
&&+ z\bigg[\frac{3}{16\pi} + 8\pi^2\left(-79C_4 + 28 C_5 -8 C_6 -1336\pi C_1^2
+704\pi C_1C_2 - 64\pi C_2^2 -128\pi C_1C_3\right)\bigg]\cr
&&+ z\ln\left(\frac{z}{8\pi^2}\right)\bigg[24C_3 + 224\pi^2 C_8 - 632\pi^2 C_7 - 64\pi^2 C_9
+\frac{509}{9}C_1-\frac{76}{3}C_2\bigg]- \frac{13711}{6912\pi^3} z\left[\ln\left(\frac{z}{8\pi^2}\right)\right]^2, \label{app_ChPT_P8}
\end{eqnarray}
\end{widetext}
where, as it partially observed in Ref.~\cite{Fu:2013ffa},
both methods share same definitions of $C_i(i=0-9)$
in Eq.~(\ref{app_ChP_C9}).
It is ready to check that both approaches arrive at pretty close numerical results, as expected.
It should be worthwhile to stress that, from the quantitative discussions in Appendix~\ref{app:R NNLO},
$\tilde{r}_i(i=1,...,6)$ values are quite different from  these two scales.

\section{The scale dependence of coupling constant $\tilde{r}_i(i=1-6)$}
\label{app:R NNLO}
The scale dependence of six coupling constant $\tilde{r}_i(i=1-6)$ is imposed
to meet the requirement $\mu \left(d\bar{b}_i/d\mu\right)\equiv d\bar{b}_i/d\ln\left(\mu\right)=0$,
which leads $\bar{b}_i$ to be scale independent.
Substituting the $\bar{b}_i$'s expressions in Eq.~(B.3) of Ref.~\cite{Colangelo:2001df}
into these requirements, we can get the expressions
for $d\tilde{r}_i\left(\mu\right)/d\ln\left(\mu\right)\,(i=1-6)$,
which are solely dependent on the scale $\mu$ and the constants $\bar{\ell}_i(i=1-4)$,
and can be elegantly integrated into:
\begin{eqnarray}%
\tilde{r}_1(\mu) &=&
\left[\frac{104}{9}\bar{\ell}_1 + \frac{112}{9}\bar{\ell}_2 - 6\bar{\ell}_3 + 2\bar{\ell}_4 +\frac{193}{27}\right]
\ln\hspace{-0.07cm}\left(\frac{\mu}{m_\pi}\right)\cr
&&- 20\ln^2\hspace{-0.07cm}\left(\frac{\mu}{m_\pi}\right) + R_1 \cr
&&{}\cr
\tilde{r}_2(\mu) &=&
-\left[\frac{68}{3}\bar{\ell}_1+\frac{248}{9}\bar{\ell}_2 -7\bar{\ell}_3 + 2\bar{\ell}_4 +\frac{556}{27}
\right]\ln\left(\frac{\mu}{m_\pi}\right)\cr
&&+\frac{407}{9}\ln^2\left(\frac{\mu}{m_\pi}\right) + R_2 \cr
&&{}\cr
\tilde{r}_3(\mu) &=&
\left[\frac{100}{9}\bar{\ell}_1+\frac{44}{3}\bar{\ell}_2 +\frac{755}{108}\right]
\ln\hspace{-0.07cm}\left(\frac{\mu}{m_\pi}\right) \cr
&&-\frac{232}{9}\ln^2\hspace{-0.07cm}\left(\frac{\mu}{m_\pi}\right) + R_3 \cr
&&{}\cr
\tilde{r}_4(\mu) &=&
-\left[\frac{2}{9}\bar{\ell}_1 + \frac{4}{9}\bar{\ell}_2 + \frac{1}{108}\right]\ln\hspace{-0.07cm}\left(\frac{\mu}{m_\pi}\right) \cr
&&+\frac{2}{3}\ln^2\hspace{-0.07cm}\left(\frac{\mu}{m_\pi}\right) + R_4 \cr
&&{}\cr
\tilde{r}_5(\mu) &=& -\left[\frac{7}{4}\bar{\ell}_1 + \frac{107}{36}\bar{\ell}_2+\frac{29}{432}\right]
\ln\hspace{-0.07cm}\left(\frac{\mu}{m_\pi}\right)\cr
&&+\frac{85}{18} \ln^2\hspace{-0.07cm}\left(\frac{\mu}{m_\pi}\right) + R_5 \cr
&&{}\cr
\tilde{r}_6(\mu) &=& -\left[\frac{5}{36}\bar{\ell}_1+\frac{25}{36}\bar{\ell}_2 +\frac{79}{432}\right]
\ln\hspace{-0.07cm}\left(\frac{\mu}{m_\pi}\right)\cr
&&+\frac{5}{6} \ln^2\hspace{-0.07cm}\left(\frac{\mu}{m_\pi}\right) + R_6 \,,
\label{app_ChPT_R}
\end{eqnarray}
where $R_i(i=1-6)$ are six integration constants,
which can be resolved with the obtained $\tilde{r}_i(i=1-6)$ values
at the selected scale: e.g., $\mu=M_\rho=0.77~{\rm GeV}$,
since the relevant published data are usually at the scale of rho mass.
Note that $\tilde{r}_i(i=1-6)$ are all quadratic in $\ln\mu$, as expected in Ref.~\cite{Colangelo:2001df}.
As a consequence, it indicates $\tilde{r}_i$ can quickly vary with the scale $\mu$,
as displayed above in Fig.~\ref{fig:cgl_NNLO_r16}.
In practice, some fit parameters (eg., $C_4$) contain $\tilde{r}_i$,
and its phenomenological value can efficiently guide one for a successful fit. 
This means that the proper $\tilde{r}_i$ values at any scale $\mu$ are very useful.
Admittedly, the analytic expressions in Eq.~({\ref{app_ChPT_R})) are in fact indicated
in Refs.~\cite{Bijnens:1997vq,Colangelo:2001df,Bijnens:2014lea},
we here just offer its explicit expressions.
Anyway, Eq.~({\ref{app_ChPT_R}) is somewhat handy and helpful.



\begin{thebibliography}{99}
\bibitem{Weinberg:1966kf} S.~Weinberg, Pion scattering lengths,
Phys.\ Rev.\ Lett.\  {\bf 17}, 616 (1966);  
S.~Weinberg, Phenomenological Lagrangians,
Physica A {\bf 96}, 327 (1979).

\bibitem{Gasser:1983yg} J.~Gasser and H.~Leutwyler,
Chiral Perturbation Theory To One Loop,
Ann. Phys.(N.Y.) \  {\bf 158}, 142 (1984). 

\bibitem{Bijnens:1995yn} J.~Bijnens, G.~Colangelo, G.~Ecker, J.~Gasser and M.~E.~Sainio,
Elastic $\pi\pi$ scattering to two loops,
Phys.\ Lett.\ B {\bf 374}, 210 (1996). 

\bibitem{Bijnens:1997vq}  J.~Bijnens, G.~Colangelo, G.~Ecker, J.~Gasser and M.~E.~Sainio,
Pion pion scattering at low energy,
Nucl.\ Phys.\ B {\bf 508}, 263 (1997)  [Erratum-ibid.\ B {\bf 517}, 639 (1998)].

\bibitem{Colangelo:2001df} G.~Colangelo, J.~Gasser and H.~Leutwyler,
$\pi\pi$ scattering, Nucl.\ Phys.\ B {\bf 603}, 125 (2001). 

\bibitem{Bijnens:2014lea} J.~Bijnens and G.~Ecker,
Mesonic low-energy constants,
Ann.\ Rev.\ Nucl.\ Part.\ Sci.\  {\bf 64}, 149 (2014).

\bibitem{Pislak:2003sv} S.~Pislak, R.~Appel, G.~S.~Atoyan, B.~Bassalleck,
D.~R.~Bergman, N.~Cheung, S.~Dhawan and H.~Do {\it et al.},
High statistics measurement of K(e4) decay properties,
Phys.\ Rev.\ D {\bf 67}, 072004 (2003). 

\bibitem{Batley:2010zza} J.~R.~Batley{\,\it et al.}(NA48-2 Collaboration),
Precise tests of low energy QCD from K(e4)decay properties,
Eur.\ Phys.\ J.\ C {\bf 70}, 635 (2010).

\bibitem{NPLQCD:2011htk} S.~R.~Beane, E.~Chang, W.~Detmold, H.W.~Lin, T.~C.~Luu,
K.~Orginos, A.~Parreno, M.~J.~Savage, A.~Torok, and A. Walker-Loud, (NPLQCD),
The I=2 $\pi\pi$ S-wave Scattering Phase Shift from Lattice QCD,
Phys. Rev. D \textbf{85}, 034505 (2012).

\bibitem{Dudek:2012gj} J.~J.~Dudek, R.~G.~Edwards and C.~E.~Thomas,
S and D-wave phase shifts in isospin-2 $\pi\pi$ scattering from lattice QCD,
Phys. Rev. D \textbf{86}, 034031 (2012).

\bibitem{Sasaki:2013vxa} 
K.~Sasaki, N.~Ishizuka, M.~Oka and T.~Yamazaki, (PACS-CS Collaboration),
Scattering lengths for two pseudoscalar meson systems,
Phys.\ Rev.\ D {\bf 89}, no. 5, 054502 (2014).

\bibitem{Helmes:2015gla}
C.~Helmes, C.~Jost, B.~Knippschild, L.~Liu, C.~Urbach, M.~Ueding, M.~Werner, C.~Liu, J.~Liu, and Z.~Wang
(ETM Collaboration), Hadron-hadron interactions from N$_{f}$ = 2 + 1 + 1 lattice QCD: isospin-2 $\pi\pi$ scattering length,
J. High Energy Phys. {\bf 1509}, 109 (2015).

\bibitem{Fischer:2020jzp}
M.~Fischer, B.~Kostrzewa, L.~Liu, F.~Romero-L\'opez, M.~Ueding and C.~Urbach,
Scattering of two and three physical pions at maximal isospin from lattice QCD,
Eur. Phys. J. C \textbf{81}, no.5, 436 (2021).

\bibitem{Sharpe:1992pp}
S.~R.~Sharpe, R.~Gupta and G.~W.~Kilcup,
Lattice calculation of I = 2 pion scattering length,
Nucl. Phys. B \textbf{383}, 309-354 (1992).

\bibitem{Kuramashi:1993ka} Y.~Kuramashi, M.~Fukugita, H.~Mino, M.~Okawa and A.~Ukawa,
Lattice QCD Calculation of Full Pion Scattering Lengths,
Phys. Rev. Lett.  {\bf 71} (1993) 2387.

\bibitem{Fukugita:1994ve} M.~Fukugita, Y.~Kuramashi, M.~Okawa, H.~Mino and A.~Ukawa,
Hadron scattering lengths in lattice QCD,
Phys. Rev. D \textbf{52}, 3003-3023 (1995).

\bibitem{CP-PACS:2002wru} S.~Aoki \textit{et al.} [CP-PACS],
I = 2 pion scattering length with Wilson fermions,
Phys. Rev. D \textbf{67}, 014502 (2003).

\bibitem{CP-PACS:2004dtj} T.~Yamazaki \textit{et al.} [CP-PACS],
I = 2 $\pi\pi$ scattering phase shift with two flavors of O(a) improved dynamical quarks,
Phys. Rev. D \textbf{70}, 074513 (2004).

\bibitem{Beane:2005rj} S.~R.~Beane, Paulo~F.~Bedaque, Kostas~Orginos, and Martin~J.~Savage (NPLQCD Collaboration),
I = 2 $\pi\pi$ scattering from fully-dynamical mixed-action lattice QCD,
Phys. Rev. D \textbf{73}, 054503 (2006).

\bibitem{Beane:2007xs}
S.~R.~Beane, T.~C.~Luu, K.~Orginos, A.~Parreno, M.~J.~Savage, A.~Torok and A.~Walker-Loud,
Precise Determination of the I=2 $\pi\pi$ Scattering Length from Mixed-Action Lattice QCD,
Phys. Rev. D \textbf{77}, 014505 (2008).

\bibitem{Sasaki:2008sv}
K.~Sasaki and N.~Ishizuka,  $I=2$ Two-Pion Wave Function and Scattering Phase Shift,
Phys. Rev. D \textbf{78}, 014511 (2008).

\bibitem{Dudek:2010ew}
J.~J.~Dudek, R.~G.~Edwards, M.~J.~Peardon, D.~G.~Richards and C.~E.~Thomas,
Phase shift of isospin-2 $\pi\pi$ scattering from lattice QCD,
Phys. Rev. D \textbf{83}, 071504(R) (2011).

\bibitem{Feng:2009ij} X.~Feng, K.~Jansen and D.~B.~Renner,
The $\pi^+\pi^+$ scattering length from maximally twisted mass lattice QCD,
Phys. Lett. B \textbf{684}, 268-274 (2010).

\bibitem{Yagi:2011jn}
T.~Yagi, S.~Hashimoto, O.~Morimatsu and M.~Ohtani,
I=2 $\pi$-$\pi$ scattering length with dynamical overlap fermion, [arXiv:1108.2970 [hep-lat]].

\bibitem{Fu:2013ffa} Z.~Fu,
Lattice QCD study of the s-wave $\pi\pi $ scattering lengths in the I=0 and 2 channels
Phys.\  Rev.\  D 87, {\bf 074501} (2013).

\bibitem{Kurth:2013tua}
T.~Kurth, N.~Ishii, T.~Doi, S.~Aoki and T.~Hatsuda,
Phase shifts in $I=2 {\pi}{\pi}$-scattering from two lattice approaches,
JHEP \textbf{12}, 015 (2013).

\bibitem{Bulava:2016mks}
J.~Bulava, B.~Fahy, B.~H\"orz, K.~J.~Juge, C.~Morningstar and C.~H.~Wong,
$I=1$ and $I=2$ $\pi-\pi$ scattering phase shifts from $N_{\mathrm{f}} = 2+1$ lattice QCD,
Nucl. Phys. B \textbf{910}, 842-867 (2016).

\bibitem{HALQCD:2017xsa}
D.~Kawai \textit{et al.} [HAL QCD],
$I=2$ $\pi\pi$ scattering phase shift from the HAL QCD method with the LapH smearing,
PTEP \textbf{2018}, no.4, 043B04 (2018).

\bibitem{RBC:2021acc} T.~Blum \textit{et al.} [RBC and UKQCD],
Lattice determination of I=0 and 2 \ensuremath{\pi}\ensuremath{\pi} scattering phase shifts with a physical pion mass,
Phys. Rev. D \textbf{104}, no.11, 114506 (2021).

\bibitem{Rodas:2023gma}
Arkaitz Rodas, Jozef J. Dudek, and Robert G. Edwards [Hadron Spectrum],
Quark mass dependence of \ensuremath{\pi}\ensuremath{\pi} scattering in isospin 0, 1, and 2 from lattice QCD,
Phys. Rev. D \textbf{108}, no.3, 034513 (2023)

\bibitem{Culver:2019qtx}
C.~Culver, M.~Mai, A.~Alexandru, M.~D\"oring and F.~X.~Lee,
Pion scattering in the isospin $I=2$ channel from elongated lattices,
Phys. Rev. D \textbf{100}, no.3, 034509 (2019).

\bibitem{Roy:1971tc} S.~M.~Roy,
Exact integral equation for pion pion scattering involving only physical region partial waves,
Phys.\ Lett.\  B {\bf 36}, 353 (1971).  

\bibitem{Basdevant:1973ru} J.~L.~Basdevant, C.~D.~Froggatt and J.~L.~Petersen,
Construction Of Phenomenological Pi Pi Amplitudes,
Nucl.\ Phys.\  B {\bf 72}, 413 (1974).  

\bibitem{Ananthanarayan:2000ht} B.~Ananthanarayan, G.~Colangelo, J.~Gasser and H.~Leutwyler,
Roy equation analysis of pi pi scattering,
Phys.\ Rept.\  {\bf 353}, 207 (2001).

\bibitem{GarciaMartin:2011cn} R.~Garcia-Martin, R.~Kaminski, J.~R.~Pelaez, J.~Ruiz de Elvira and F.~J.~Yndurain,
The Pion-pion scattering amplitude. IV: Improved analysis with once subtracted Roy-like equations up to 1100 MeV,
Phys.\ Rev.\  {\bf D 83}, 074004 (2011). 

\bibitem{Fu:2017apw} Z.~Fu and X.~Chen,
$I=0$ $\pi\pi$ $s$-wave scattering length from lattice QCD,
Phys.\ Rev.\ D {\bf 98}, 014514 (2018)

\bibitem{Beane:2003da} S.~R.~Beane, P.~F.~Bedaque, A.~Parre\~no and M.~J.~Savage,
Two nucleons on a lattice,
Phys.\ Lett.\  B {\bf 585}, 106 (2004).

\bibitem{Adhikari:1983ii} S.~K.~Adhikari and J.~R.~A.~Torreao,
Effective Range Expansion For The Pion-Pion System,
Phys.\ Lett.\  {\bf 123B}, 452 (1983).  

\bibitem{Hanhart:2008mx} C.~Hanhart, J.~R.~Pelaez and G.~Rios,
Quark mass dependence of the rho and sigma from dispersion relations
and Chiral Perturbation Theory,
Phys.\ Rev.\ Lett.\  {\bf 100}, 152001 (2008).

\bibitem{Hanhart:2014ssa} C.~Hanhart, J.~R.~Pelaez and G.~Rios,
Remarks on pole trajectories for resonances,
Phys.\ Lett.\ B {\bf 739}, 375 (2014).

\bibitem{Albaladejo:2012te} M.~Albaladejo and J.~A.~Oller,
On the size of the sigma meson and its nature,
Phys.\ Rev.\ D {\bf 86}, 034003 (2012).

\bibitem{Pelaez:2010fj} J.~R.~Pelaez and G.~Rios,
Chiral extrapolation of light resonances from one and two-loop unitarized Chiral Perturbation Theory versus lattice results,
Phys.\ Rev.\ D {\bf 82}, 114002 (2010).

\bibitem{Bernard:2001av} C. Bernard, T. Burch, K. Orginos, D. Toussaint,
T. A. DeGrand, C. DeTar, S. Datta, S. Gottlieb, U. M. Heller, and R. Sugar,
The QCD spectrum with three quark flavors,
Phys.\ Rev.\ D {\bf 64}, 054506 (2001).

\bibitem{Aubin:2004wf} 
C. Aubin, C. Bernard, C. DeTar, J. Osborn, S. Gottlieb, E. B.
Gregory, D. Toussaint, U. M. Heller, J. E. Hetrick, and R. Sugar,
Light hadrons with improved staggered quarks: Approaching the continuum limit,''
Phys.\ Rev.\ D {\bf 70}, 094505 (2004).

\bibitem{stag_fermion}
K.~Orginos, D.~Toussaint and R.~L.~Sugar,
Variants of fattening and flavor symmetry restoration,
Phys.\ Rev.\  D {\bf 60}, 054503 (1999); 
T. Blum, C. DeTar, S. Gottlieb, K. Rummukainen, U.M. Heller, J. E. Hetrick, D. Toussaint,
R. L. Sugar, and M. Wingate,
Improving flavor symmetry in the Kogut-Susskind hadron spectrum,
Phys.\ Rev.\ D {\bf 55}, R1133 (1997); 
G.~P.~Lepage,  Flavor-symmetry restoration and Symanzik improvement for staggered quarks,
Phys.\ Rev.\ D {\bf 59}, 074502 (1999); 
C. Bernard, T. Burch, T. A. DeGrand, C. DeTar, S. Gottlieb, U. M. Heller ,
J. E. Hetrick, K. Orginos, B. Sugar and D. Toussaint,
 Scaling tests of the improved Kogut-Susskind quark action,
Phys.\ Rev.\ D {\bf 61}, 111502(R) (2000). 

\bibitem{Luscher:1986pf} M.~L\"uscher,
Volume Dependence of the Energy Spectrum in Massive Quantum Field Theories. 2. Scattering States,
Commun.\ Math.\ Phys.\  {\bf 105} (1986) 153.

\bibitem{Luscher:1990ux} M.~L\"uscher,
Two particle states on a torus and their relation to the scattering matrix,
Nucl.\ Phys.\  {\bf B 354}, 531 (1991).

\bibitem{Luscher:1990ck} M.~L\"uscher, U.~Wolff,
How to Calculate the Elastic Scattering Matrix in Two-dimensional Quantum Field Theories by Numerical Simulation,
Nucl.\ Phys.\ B {\bf 339}, 222 (1990).

\bibitem{Rummukainen:1995vs} K.~Rummukainen and S.~A.~Gottlieb,
Resonance scattering phase shifts on a nonrest frame lattice,
Nucl.\ Phys.\  B {\bf 450}, 397 (1995). 

\bibitem{Kim:2005gf} C.~h.~Kim, C.~T.~Sachrajda and S.~R.~Sharpe,
Finite-volume effects for two-hadron states in moving frames,
Nucl.\ Phys.\  B {\bf 727}, 218 (2005).

\bibitem{Christ:2005gi} N.~H.~Christ, C.~Kim and T.~Yamazaki,
Finite volume corrections to the two-particle decay of states with non-zero momentum,
Phys.\ Rev.\  D {\bf 72}, 114506 (2005).

\bibitem{Doring:2012eu} M.~Doring, U.~G.~Meissner, E.~Oset and A.~Rusetsky,
Scalar mesons moving in a finite volume and the role of partial wave mixing,
Eur.\ Phys.\ J.\ A {\bf 48}, 114 (2012).

\bibitem{Fu:2011xz} Z.~Fu,
Rummukainen-Gottlieb's formula on two-particle system with different mass,
Phys.\ Rev.\ D {\bf 85}, 014506 (2012).

\bibitem{Leskovec:2012gb} L.~Leskovec and S.~Prelovsek,
Scattering phase shifts for two particles of different mass and non-zero total momentum in lattice QCD,
Phys.\ Rev.\ D {\bf 85}, 114507 (2012).

\bibitem{Lepage:1989hd} G. P. Lepage,
in Proceedings of TASI'89 Summer School, edited by T. DeGrand and D. Toussaint (World
Scientific, Singapore, 1990), p. 97.
The Analysis Of Algorithms For Lattice Field Theory, CLNS-89-971.

\bibitem{Fu:2016itp} Z.~Fu and L.~Wang,
Studying the $\rho$ resonance parameters with staggered fermions,
Phys.\ Rev.\ D {\bf 94}, no. 3, 034505 (2016).

\bibitem{Yamazaki:2004qb}  
T.~Yamazaki, S.~Aoki, M. Fukugita, K.-I.~Ishikawa, N.~Ishizuka, Y.~Iwasaki, K.~Kanaya, T.~Kaneko, Y.~Kuramashi, M.~Okawa, A.~Ukawa and T.~Yoshie, 
$I = 2$ $\pi\pi$ scattering phase shift with two flavors of O(a) improved dynamical quarks,
Phys.\ Rev.\ D {\bf 70}, 074513 (2004).

\bibitem{Doring:2011vk} M.~Doring, U.~-G.~Meissner, E.~Oset and A.~Rusetsky,
Unitarized Chiral Perturbation Theory in a finite volume: Scalar meson sector,
Eur.\ Phys.\ J.\ A {\bf 47}, 139 (2011). 

\bibitem{Fu:2012gf} Z.~Fu,
Preliminary lattice study of $\sigma$ meson decay width,
J. High Energy Phys. 07 (2012) 142.

\bibitem{Bernard:2010fr}  
C. Bernard, C. DeTar, M. Di Pierro, A. X. El-Khadra, R. T. Evans, E. D. Freeland, E. Gamiz, Steven Gottlieb, U. M. Heller, J. E. Hetrick,
A. S. Kronfeld, J. Laiho, L. Levkova, P. B. Mackenzie, J. N. Simone, R. Sugar, D. Toussaint, and R. S. Van de Water (Fermilab Lattice and MILC Collaborations),
Tuning Fermilab Heavy Quarks in 2+1 Flavor Lattice QCD with Application to Hyperfine Splittings,
Phys.\ Rev.\ D {\bf 83}, 034503 (2011).

\bibitem{Bazavov:2009bb}  A.~Bazavov {\it et al.} [MILC Collaboration],
Nonperturbative QCD Simulations with 2+1 Flavors of Improved Staggered Quarks,
Rev.\ Mod.\ Phys.\  {\bf 82}, 1349 (2010).

\bibitem{Kaplan:1992bt}
D.~B.~Kaplan,
A Method for simulating chiral fermions on the lattice,
Phys.\ Lett.\ B {\bf 288}, 342 (1992);
Y.~Shamir,
Reducing chiral symmetry violations in lattice QCD with domain wall fermions,
Phys.\ Rev.\ D {\bf 59}, 054506 (1999).

\bibitem{MILC:DeTar}
https://web.physics.utah.edu/$\tilde{}$detar/milc/

\bibitem{Fu:2011xw} Z.~Fu,
Preliminary lattice study of $\kappa$ meson decay width,
J. High Energy Phys. 01 (2012) 017.

\bibitem{Fu:2012tj}Z.~Fu and K.~Fu,
Lattice QCD study on $K^\ast(892)$ meson decay width,
Phys.\ Rev.\  D {\bf 86}, 094507 (2012).

\bibitem{Fu:2011wc} Z.~Fu,
Lattice study on $\pi K $ scattering with moving wall source,
Phys.\ Rev.\ D {\bf 85}, 074501 (2012);
Z.~Fu,
Preliminary lattice study of the I=1 $K \bar{K}$ scattering length,
Eur.\ Phys.\ J.\ C {\bf 72}, 2159 (2012).

\bibitem{Umeda:2007hy} T.~Umeda,
A Constant contribution in meson correlators at finite temperature,
Phys.\ Rev.\ D {\bf 75}, 094502 (2007).

\bibitem{Gupta:1993rn} R.~Gupta, A.~Patel and S.~R.~Sharpe,
I = 2 pion scattering amplitude with Wilson fermions,
Phys.\ Rev.\ D {\bf 48}, 388 (1993).

\bibitem{Golterman:1985dz} M.~F.~L.~Golterman,
Staggered Mesons,
Nucl.\ Phys.\ B {\bf 273}, 663 (1986).

\bibitem{DeTar:2014gla} C.~DeTar and S.~H.~Lee,
Variational method with staggered fermions,
Phys.\ Rev.\ D {\bf 91}, 034504 (2015). 

\bibitem{Lepage:2001ym}
G.~P.~Lepage, B.~Clark, C.~T.~H.~Davies, K.~Hornbostel,
P. B. Mackenzie, C. Morningstar, and H.~Trottier,
Constrained curve fitting,
Nucl. Phys. B Proc. Suppl. \textbf{106-107}, 12-20 (2002);
D. S. Sivia, Data Analysis: A Bayesian
Tutorial (Oxford University Press, NewYork, 1996).

\bibitem{Hu:2017wli}
B.~Hu, R.~Molina, M.~D\"oring, M.~Mai and A.~Alexandru,
Chiral extrapolations of the $\boldsymbol{\rho(770)}$ meson in $\mathbf{N_f=2+1}$ lattice QCD simulations,
Phys.\ Rev.\ D {\bf 96}, no. 3, 034520 (2017).


\bibitem{Workman:2022ynf}
R.~L.~Workman \textit{et al.} [Particle Data Group], Review of Particle Physics,
PTEP \textbf{2022}, 083C01 (2022).


\bibitem{Caprini:2011ky} I.~Caprini, G.~Colangelo and H.~Leutwyler,
Regge analysis of the $\pi\pi$ scattering amplitude,
Eur. Phys. J. C \textbf{72}, 1860 (2012).


\bibitem{Buchoff:2008ve} M.~I.~Buchoff,
Isotropic and Anisotropic Lattice Spacing Corrections for I=2 $\pi\pi$ Scattering from Effective Field Theory,
Phys. Rev. D \textbf{77}, 114502 (2008).

\bibitem{Buchoff:2008hh}
M.~I.~Buchoff, J.~W.~Chen and A.~Walker-Loud,
pi-pi Scattering in Twisted Mass Chiral Perturbation Theory,
Phys. Rev. D \textbf{79}, 074503 (2009).
\end{thebibliography}
\end{document}